\RequirePackage{fix-cm}
\documentclass[smallcondensed]{svjour3}     
\smartqed  
\usepackage{graphicx}
\usepackage{amssymb,amsmath,color,bm}
\begin{document}

\title{Refinement on non-hydrostatic shallow granular flow model in a global Cartesian coordinate system 
}


\titlerunning{Refinement on non-hydrostatic shallow granular  flow model}        

\author{L. Yuan     \and W. Liu \and J. Zhai \and S.F. Wu \and A.K. Patra \and E.B. Pitman
}


\institute{Li Yuan (corresponding author), Wei Liu, Jian Zhai   \at
             LSEC and UCAS,
Institute of Computational Mathematics and Scientific/Engineering Computing, Academy of Mathematics and Systems Science,
Chinese Academy of Sciences, Beijing 100190, CHINA. \\
                   \email{lyuan@lsec.cc.ac.cm, zhaijian@lsec.cc.ac.cn, liuwei@lsec.cc.ac.cn}           
           \and
           Shifeng Wu   \at
              Department of Computer Science, Guangdong Polytechnic Normal University, Guangzhou
510665, CHINA   \\
\email{fengtree@126.com}  \and
A.K. Patra  \at
Department of Mechanical and Aerospace Engineering, University at Buffalo, SUNY, Buffalo, NY 14260, USA  \\
\email{abani@eng.buffalo.edu} \and
E.B. Pitman \at
Department of Mathematics, University at Buffalo, SUNY, Buffalo, NY 14260, USA \\
\email{pitman@buffalo.edu}
}

\date{Received: date / Accepted: date}

\maketitle

\begin{abstract}
Current shallow granular flow models suited to arbitrary topography can {be divided} into two types, those formulated in bed-fitted curvilinear coordinates, and those formulated in global Cartesian coordinates. The shallow granular flow model of Denlinger and Iverson \cite{Denlinger2004} and the Boussinesq-type shallow granular flow theory of Castro-Orgaz \emph{et al}.  \cite{Castro2014} are formulated in a Cartesian coordinate system (with $z$ vertical), and both account for the effect of nonzero vertical acceleration  on depth-averaged momentum fluxes and stress states. In this paper, we first reformulate the vertical normal stress of Castro-Orgaz \emph{et al}. \cite{Castro2014} in a quadratic polynomial in the relative elevation $\eta$. This form allows for analytical depth integration of the vertical normal stress. We then calculate the basal normal stress based on the basal friction law and scaling analysis. These calculations, plus certain constitutive relations, lead to a refined full non-hydrostatic shallow granular flow model, which is further rewritten in a form of Boussinesq-type water wave equations for future numerical studies. In the present numerical study, we apply the open-source code TITAN2D to numerical solution of a low-order version of the
full model involving only a mean vertical acceleration correction term. To cure the numerical instability associated with discretization of the enhanced gravity,  we propose an approximate formula for the enhanced gravity by utilizing the hydrostatic pressure assumption in the bed normal direction. Numerical calculations are conducted for several test cases involving steep slopes. Comparison with a  bed-fitted model shows that even the simplified non-hydrostatic  Cartesian model can be used to simulate shallow granular flows over arbitrary topography.

\end{abstract}

\section{Introduction}
\label{intro}

Reliable prediction of
gravity-driven {geophysical} mass movements like landslides, debris flows, and rock avalanches can be an invaluable tool in assessing hazard risks and planning strategies for hazard mitigation. It is widely recognized that the basic ingredients in {geophysical} mass movements in {natural hazards} are granular materials, {a collection of a larger number of discrete solid particles with interstices filled with a  fluid or gas} \cite{Pud2007}, thus granular avalanche flows constitute a fundamental case. The relative simplicity of this case makes it an attractive object for developing and testing various predictive models \cite{Denlinger2004}.

During a granular avalanche, granular materials slide and flow over topographies and may travel very long distances. The characteristic length
in the flowing direction is generally much larger than that in the normal-to-bed direction, e.g., the avalanche thickness. Such a shallowness argument, which originated in the derivation of Saint-Venant equations for modeling shallow water flows, has been widely used in the derivation of continuum flow models for granular avalanches. Earlier shallow granular flow  models were formulated by direct analogy with shallow water equations \cite{Grigo1967}. Later, Savage and Hutter \cite{Savage1989} introduced a depth-integrated theory obeying Coulomb-type yield by
which the rapid flow of a finite mass of granular material down a plane slope could be analysed. Their shallow granular flow model is called the Savage-Hutter (SH) equations. Over the past three decades there has been great progress in shallow granular flow models. The developed models have been
shown to be able to reproduce the basic features of both experimental
dense granular flows along inclined planes with appropriate constitutive relations \cite{{Iverson1997},{Wieland1999},{Denlinger2001},{Pouli2002},{Midi2004},{Mangey2007},{Mae2013}},  and some of which have been used to simulate real avalanche
flows over natural terrains \cite{{McD2008},{Luca2009},{Kuo2009}}.

In describing debris flows over natural terrains, {some} researchers \cite{{Taka1992},{Wu2013}} used the Saint-Venant equations that is referenced in horizontal Cartesian coordinates with a hydrostatic {basal} pressure of $\rho gh$   (where $\rho$ is the {bulk} density, $h$ is the vertical {flow} depth, and $g$ is the gravity of the Earth). However, such global Cartesian formulations are only applicable to topography with small slopes because the usual hypothesis of hydrostatic pressure in the vertical direction in the shallow water equations is no longer valid for steep terrain even if it is admissible in the normal-to-bed direction.
A few studies \cite{{Patra2005},{Chau2004}} directly used formulations in a local Cartesian coordinate system in cell-by-cell way to compute granular flows over natural terrains.  However, such a numerical approach is problematic in aligning velocity variables and balancing numerical fluxes of conserved variables between adjacent cells on curved bed as remarked by Denlinger and Iverson \cite{Denlinger2004}. A more elaborate work was to correct both acceleration and friction terms in the Saint-Venant equations in Cartesian coordinates for large slope gradients \cite{Herg2015}. Nevertheless, the {corrections are based on mechanical considerations  rather than on mathematical derivations}.  On the other hand, the SH theory has been generalized rigorously in general curvilinear coordinate systems to describe granular avalanches over general terrains \cite{{Gray1999},{Pud2003},{Luca2009}}.
Although the bed-fitted formulations are more accurate,  they are complicated and need non-trivial grid generation on natural terrains for numerical solution.
In order to develop viable shallow water/granular flow models suitable for a general topography, Refs.
\cite{{Bouchut2003},{Bouchut2004}} derived a form of shallow water/granular flow equations that is referenced to a fixed global Cartesian coordinate system but uses thickness in the direction normal to the topography and  a parameterized Cartesian velocity field as solution variables. The equations \cite{Bouchut2004} take into account the curvature tensor with all its components and the Coulomb basal friction  while ignoring the internal friction effects.
Numerical solution using this model has been carried out for landslide over general terrains \cite{Kuo2009}.

For modeling gravity-driven granular avalanche flows across irregular terrains, Denlinger and Iverson
\cite{Denlinger2004} developed depth-averaged governing equations in a global Cartesian coordinate system (with $z$ vertical) that account explicitly for the effect of nonzero vertical accelerations on depth-averaged momentum fluxes and stress states. They used stress transformation between the bed-fitted local Cartesian and the horizontal global Cartesian coordinate systems to get the Coulomb stress states independent of the orientation of
the coordinate system. While this model provides familiar conservative fluxes suitable for finite volume methods, the source terms contain internal stresses which are calculated with finite element methods, and this will introduce extra work.

More recently,  Castro-Orgaz \emph{et al}. \cite{Castro2014} developed a non-hydrostatic depth-averaged granular flow theory in the global horizontal-vertical Cartesian coordinate system
by making use of the non-hydrostatic Boussinesq-type water wave theory widely used in water wave field  (e.g., \cite{{Green-Naghdi76}}). In their theory, the effect of vertical motion is taken into account rigorously, and the vertical velocity, vertical acceleration, and vertical normal stresses are determined mathematically using mass and momentum conservation equations without any \emph{ad-hoc}  simplification. Some basic features of this theory were explored and analytical solutions of simplified flow cases were obtained, and numerical approaches  for treating the additional dispersive terms in the fluxes of the depth-averaged momentum equations were outlined \cite{Castro2014}. In almost the same time, Ref. \cite{Brist2015} {also} derived a non-hydrostatic shallow water-type model by a minimal energy constraint and depth-averaging process of the Euler or Navier-Stokes system, and  the resulting model is similar to the non-hydrostatic shallow granular flow theory \cite{Castro2014}  in a sense that a
non-hydrostatic part of the pressure will be determined using additional equations.

In this paper, we further develop the non-hydrostatic depth-averaged granular flow theory \cite{Castro2014}. First, we reformulate the vertical normal stress  \cite{Castro2014} in a quadratic polynomial form in the relative elevation $\eta=z-b(x,y,t)$ ($b$ is the topography). This new form is convenient for analytical depth integration, and it also reveals the essential difference between the model \cite{Denlinger2004} and the theory \cite{Castro2014}.  Second, we compute the basal traction vector from the Coulomb friction law as Refs. \cite{{Gray2003},{Mangey2003}} did. {In this step}, we find the vertical component of the basal traction vector is linked to the integration of the $z-$momentum equation such that an accurate expression for the basal normal stress in terms of the {enhanced gravity} can be obtained under some scaling {arguments}. With the {above two revisions},  a refined complete non-hydrostatic shallow granular flow model is obtained under certain constitutive relations, in which the earth pressure coefficient {notation for the lateral normal stresses in soil mechanics} and the relation between the lateral shear and normal stresses  \cite{Denlinger2001} are adopted. These relations can be replaced or improved in future work.  The resultant full model is further rewritten in a form of Boussinesq-type water wave equations presumedly more suitable for applying mature numerical methods developed by the water wave community.

Since the present full non-hydrostatic shallow granular flow equations are still complicated, we only carry out numerical solution of its low order version, which is similar to the differential form of the model \cite{Denlinger2004}. We implement the lower order model in the open-source shallow granular flow simulation code TITAN2D \cite{titancode}. In the implementation, we encountered numerical instability problem caused by discretizing $\partial \bar{w}/\partial t$ in the enhanced gravity defined as  $g^\prime=g+D\bar{w}/Dt$ . To overcome this problem, we derive an approximate formula for $g^\prime$ by using the hypothesis of hydrostatic pressure in the bed normal direction and the Taylor expansion. This formula {takes into account the effects of bed slope, basal friction, and variation of flow height}, and is  found to be numerically more stable than  {the original enhanced gravity}. In addition, a more delicate  ``centripetal normal stress" by using the curvature tensor \cite{Mangey2007} is added to the basal normal stress in the basal friction and bed slope terms.
The resultant simplified non-hydrostatic model is implemented in TITAN2D and tested in several numerical examples. Numerical comparisons with analytical solutions and bed-fitted model show a satisfactory agreement.

\section{A Survey of Two Non-hydrostatic Models}\label{sec:1}

For later reference in this paper, we briefly review the basic equations of granular
avalanche motions and the shallow  granular flow equations of Denlinger and Iverson \cite{Denlinger2004} and Castro-Orgaz \emph{et al}. \cite{Castro2014},  respectively.

\subsection{Conservation equations}

\begin{figure}[hp!]
\centering
\includegraphics[width=16pc]{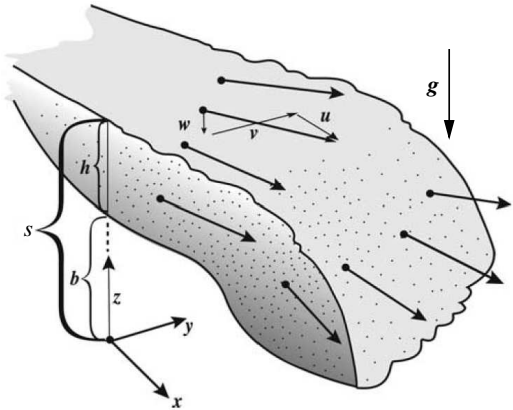}
\caption{Sketch illustrating the global coordinate
system and variables used to calculate avalanche motion (courtesy of Denlinger and Iverson \cite{Denlinger2004}).}
\label{figure1}
\end{figure}

In a  horizontal-vertical Cartesian  coordinate system  where the $z$ direction is opposite to the gravitational acceleration vector $\mathbf{g}$  (figure \ref{figure1}),
the motion of a fluidized granular mass is described with the mass and momentum conservation equations
\begin{eqnarray}
 \mbox{div~} \mathbf{v} &= & 0   \label{cont}  ,\\
\rho \left[ \frac{\partial \mathbf{v}}{\partial t}  + \mbox{div}\left( \mathbf{v}\otimes\mathbf{v}\right)\right] & =& -\mbox{div~}\bm{\tau} +\rho  \mathbf{g}  , \label{mom}
\end{eqnarray}
where $t$ is the time, $\rho$ is the
bulk density of the granular mass  assumed to be constant here,  $\mathbf{v}(x, y, z, t)  = (u(x, y, z, t),  v(x, y, z, t),   w(x, y, z, t)) $
denotes the 3D velocity vector inside the
avalanche, $\otimes$  is the tensor (or dyadic) product, $\bm{\tau}(x, y, z, t)$ is the pressure tensor (the negative Cauchy stress).

Kinematic boundary conditions are imposed on the free surface $z=s(x,y,t)$ and the basal surface $z=b(x,y,t)$, that specify that mass neither enters nor leaves at the free surface or at the base:
\begin{eqnarray}
\label{kine1}
&& \left.\left( \frac{\partial s}{\partial t} + u\frac{\partial s}{\partial x}+v\frac{\partial s}{\partial y} -w \right)\right|_{z=s} =0  , \\
&& \left.\left(\frac{\partial b}{\partial t}+ u\frac{\partial b}{\partial x}+v\frac{\partial b}{\partial y} -w \right)\right|_{z=b} =0.
\label{kine2}
\end{eqnarray}
The dynamic boundary conditions include  a traction-free boundary condition at the free surface,
and a Coulomb sliding friction law at the basal surface \cite{Gray2003}:
\begin{eqnarray}
&&\bm{\tau}_s\cdot \mathbf{n}_s= \mathbf{0}   , \label{dy1} \\
&&\bm{\tau}_b \cdot \mathbf{n}_b= \frac{\mathbf{v}_r}{|\mathbf{v}_r|} \tan \phi_\text{bed} (\mathbf{n}_b\cdot  \bm{\tau}_b\cdot \mathbf{n}_b)+\mathbf{n}_b(\mathbf{n}_b\cdot \bm{\tau}_b\cdot \mathbf{n}_b),
\label{dy2}
\end{eqnarray}
where the outward unit normals (to the outside of the granular mass) are defined as  $\mathbf{n}_s=(-\partial_x s, - \partial_y s, 1)/\sqrt{1+(\partial_x s)^2+(\partial_y s)^2}$ and  $\mathbf{n}_b= (\partial_x b, \partial_y b, -1)$ $ /\sqrt{1+(\partial_x b)^2+(\partial_y b)^2}$, respectively, $\phi_\text{bed}$ is the basal angle of friction and $\mathbf{v}_r=\mathbf{v}_{b+}-\mathbf{v}_{b-} $ is the velocity
difference (satisfying $\mathbf{v}_r\cdot \mathbf{n}_b=0$)  between the fluid on the upper side of the basal surface, $\mathbf{v}_{b+}$, and the basal topography on the lower side of the interface,  $\mathbf{v}_{b-}$.  The factor $\mathbf{v}_r/|\mathbf{v}_r|$ ensures that the Coulomb friction opposes the avalanche motion. For a fixed bed, $\mathbf{v}_{b-}=0$.

\subsection{Shallow granular flow model of Denlinger and Iverson (2004)}

Denlinger and Iverson
\cite{Denlinger2004} derived a depth-averaged granular flow model in the global Cartesian coordinate system as shown in figure \ref{figure1}.
They started from conservation equations (\ref{cont})-(\ref{mom}) and boundary conditions (\ref{kine1})-(\ref{dy1}) and noted that the length scale for avalanche thickness in the $z$ direction is $H$, whereas the length scale for typical planimetric zone of the avalanche in the $x$ and $y$ directions is $L$.  They assumed  $H\ll L$ typically so that the parameter $\varepsilon=H/L $ is much less than unity.
By scaling considerations and integration of the $z$-component momentum equation across the avalanche  thickness with a  stress free condition $\tau_{zz}(s)=0$ at the free surface, they obtained the vertical normal stress $\tau_{zz}(b)$ at the bed in terms of a hydrostatic pressure plus a depth-averaged vertical acceleration {correction} term. Then they assumed that $\tau_{zz}(z)$ varies
 linearly from $\tau_{zz}(b)$ at the basal surface to 0 at the free surface and that the lateral normal stresses $\tau_{xx}$ and $\tau_{yy}$ are proportional to $\tau_{zz}$, and assumed a constant velocity profile for the horizontal velocity components $(u,v)$ across the vertical thickness.  With these assumptions, they derived following  depth-averaged mass and horizontal momentum equations (we correct the signs and the bed slope term typos in \cite{Denlinger2004} in the right-hand side (RHS) of equations (\ref{x-mom}) and (\ref{y-mom}))
\begin{eqnarray}
 \int_A \left[
\frac{\partial h}{\partial t} + \frac{\partial( h\bar{u})}{\partial x} +\frac{\partial(h\bar{v})}{\partial y} \right] \mbox{d} A  &= & 0, \label{depth} \\
\int_A \left[\frac{\partial (h \bar{u})}{\partial t} + \frac{\partial}{\partial x} \left( h\bar{u}^2+\frac 12 k_x g^\prime h^2 \right)  +
\frac{\partial (h \bar{u}\bar{v} )}{\partial y} \right] \mbox{d} A&=&-\int_A k_x g^\prime h \frac{\partial b}{\partial x} \mbox{d} A \label{x-mom}  \\
&&-\int_V \frac{\partial \tau_{yx}}{\partial y}\mbox{d}V + \int_A \tau_{zx}(b) \mbox{d} A, \nonumber \\
\int_A \left[\frac{\partial (h \bar{v})}{\partial t} + \frac{\partial( h \bar{u}\bar{v}) }{\partial x} +
\frac{\partial}{\partial y}\left( h \bar{v}^2+\frac 12 k_y g^\prime h^2\right) \right] \mbox{d} A &=&-\int_A k_y g^\prime h \frac{\partial b}{\partial y} \mbox{d} A    \label{y-mom} \\
&&- \int_V \frac{\partial \tau_{xy}}{\partial x}\mbox{d}V + \int_A \tau_{zy}(b) \mbox{d} A, \nonumber
\end{eqnarray}
where $V$ is an arbitrary control volume, $A$ is its projected area on the horizontal $Oxy$ plane, $h=s-b$ is the flow thickness, measured vertically
from the bed at $z=b$ to the free surface at $z=s$, $\bar{u},\bar{v}$ and $\bar{w}$ are
velocity components in the $x, y$, and $z$ directions, averaged over the vertical thickness $h$ like  $\bar{u}= \int_b^s u\mbox{d} \zeta/h $ where $\zeta$ is the dummy variable for vertical integration,
$k_x=\tau_{xx}/\tau_{zz}$ and $k_y=\tau_{yy}/\tau_{zz}$ are lateral normal stress coefficients that have
values directly derived from Coulomb stress calculations and are independent of the orientation of
the coordinate system,  $\tau_{ij}$ are Cartesian components of the stress tensor, and $g^\prime$ is the ``total vertical acceleration" \cite{Denlinger2004} ({we adopt the term ``enhanced gravity" \cite{Castro2014} as additional terms due to the vertical acceleration are not included} in $g^\prime$) defined  by
\begin{eqnarray}
g^\prime&\equiv& g+\frac{D \bar{w}}{D t},  \label{acce}  \\
\frac{D\bar{w}}{D t}&\equiv&\frac{\partial \bar{w}}{\partial t}   + \bar{u}\frac{\partial \bar{w}}{\partial x}
+ \bar{v} \frac{\partial \bar{w}}{\partial y},  \label{dwdt}\\
\label{wav}
\bar{w} &=& \frac 12 \left(w_s+w_b\right)=  \left(\frac{\partial b}{\partial t}+ \bar{u}\frac{\partial b}{\partial x}+ \bar{v}\frac{\partial b}{\partial y} \right)+\frac 12\left(\frac{\partial h}{\partial t} + \bar{u}\frac{\partial h}{\partial x}+ \bar{v}\frac{\partial h}{\partial y}\right).
\end{eqnarray}
The governing equations (\ref{depth})--(\ref{y-mom}) are closed by equations (\ref{acce})--(\ref{wav}). Denlinger and Iverson \cite{Denlinger2004} solved the equations with finite volume methods using stresses from the previous time step in the RHS source terms.  Once a flow solution was obtained, a finite element method was used to calculate
internal stresses with dynamic boundary conditions (\ref{dy1})-(\ref{dy2}) and to modify these source terms for the next time step.

We will show in Sec.\text{~}\ref{sec:reformu} that  the differential form of equations (\ref{depth})--(\ref{y-mom}) plus (\ref{acce})--(\ref{wav})
is a low-order version of a full model refined from the following non-hydrostatic shallow granular flow theory \cite{Castro2014}. And a substitute for $g^\prime$ seems to be necessary for curing the numerical instability in discretizing equation (\ref{acce}).

\subsection{Non-hydrostatic shallow granular flow theory of Castro-Orgaz (2014)}\label{sec:castro}
Castro-Orgaz \emph{et al}. \cite{Castro2014} derived a non-hydrostatic Boussinesq-type gravity wave theory for granular media in the same Cartesian coordinate system as shown in figure \ref{figure1}. They adopted same {assumptions of shallowness for the vertical depth and constant velocity profile} for the horizontal velocity components $(u,v)$ across the vertical thickness \cite{Denlinger2004}. Starting from equations (\ref{cont})-(\ref{dy1}), they derived  the following governing equations
\begin{eqnarray}
&&\frac{\partial h}{\partial t} + \frac{\partial( h\bar{u})}{\partial x} +\frac{\partial(h\bar{v})}{\partial y}= 0, \label{depth2} \\
&&\frac{\partial (h \bar{u})}{\partial t} + \frac{\partial\left(\displaystyle h \bar{u}^2 +\frac{h\bar{\tau}_{xx}}{\rho}  \right) }{\partial x} +
\frac{\partial \left(\displaystyle h \bar{u}\bar{v} +\frac{h\bar{\tau}_{yx}}{\rho}\right)}{\partial y}  = -\frac{1}{\rho}\left(\tau_{xx}\frac{\partial b}{\partial x}+\tau_{yx}\frac{\partial b}{\partial y}-\tau_{zx}\right)_b, \nonumber  \\
\label{x-mom2}  \\
&&\frac{\partial (h \bar{v})}{\partial t} + \frac{\partial\left(\displaystyle h \bar{u}\bar{v}+ \frac{h \bar{\tau}_{xy}}{\rho} \right) }{\partial x} +
\frac{\partial \left(h\displaystyle \bar{v}^2 +\frac{h  \bar{\tau}_{yy}}{\rho} \right)}{\partial y}  = -\frac{1}{\rho}\left(\tau_{xy}\frac{\partial b}{\partial x}+\tau_{yy}\frac{\partial b}{\partial y}-\tau_{zy}\right)_b, \nonumber \\
\label{y-mom2}
&& \tau_{zz} =\rho g (h-\eta) + \rho \left[ \frac{\partial I }{\partial t}+ \nabla \cdot (\bar{\mathbf{u}} I)\right] - \rho w^2 \label{tauzz_0}, \\
&& I\equiv\int^s_z w \mbox{d} \zeta =(h-\eta)\frac{\partial b}{\partial t} -\nabla  \cdot \left [\frac {\left(h^2-\eta^2\right)}{2} \bar{\mathbf{u}} \right] +h \bar{\mathbf{u}} \cdot \nabla (h+b),  \label{Ifinal}\\
&& w= \displaystyle w_b  -(\nabla\cdot  \bar{\mathbf{u}}) \eta  , \label{wlinear2}
\end{eqnarray}
where  $\eta=z-b, \nabla =(\partial_x, \partial_y)$,   $\bar{\mathbf{u}}=(\bar{u},\bar{v})$, and
a quantity with  bar is the depth-averaged quantity.
Note that equations (\ref{depth2}), (\ref{x-mom2}) and (\ref{y-mom2}) are usual depth-averaged mass and horizontal momentum equations. But equation (\ref{tauzz_0}) results from integration of the $z-$momentum equation from a generic elevation $z$ to the free surface $z=s$  where a stress-free condition $\tau_{zz}(s)=0$ is used (the same as \cite{Denlinger2004}), {and constant profile of $\bar{\mathbf{u}}$ and negligence of shear stress contributions in equation (\ref{tauzz_0}) are implied}. Equation (\ref{Ifinal}) is a definition of $I$ and is calculated from vertical velocity component $w$ given in (\ref{wlinear2}),  which results from integration of the continuity equation (\ref{cont}) from the bed with the kinematic boundary condition (\ref{kine2}) to {a generic elevation $z$}.

The system of equations (\ref{depth2})-(\ref{wlinear2}) are closed if parameterizations of  the stress tensor $\bm \tau$ are given. Equations (\ref{tauzz_0}), (\ref{Ifinal}), and (\ref{wlinear2}) are said to be the core for modeling dispersion effects in depth-averaged models \cite{Castro2014}.  Castro-Orgaz \emph{et al}. \cite{Castro2014} compared their theory with the Denlinger and Iverson model \cite{Denlinger2004} in 1D steady dry granular flow over a horizontal plane and shown that the latter model introduces a factor
(1/4) into the dispersive terms {in the momentum flux} as compared with the exact factor (1/3) in their full non-hydrostatic shallow granular flow theory.  As for numerical solution of the full non-hydrostatic shallow granular flow equations, they mentioned numerical difficulties introduced by dispersion terms and suggested some solution methods developed in water wave simulations.

We remark that integral forms (\ref{depth})--(\ref{y-mom}) can be transformed into differential forms (\ref{depth2})--(\ref{y-mom2}) by removing the surface integral, utilizing  $\tau_{xx}=k_x\tau_{zz},\tau_{yy}=k_y\tau_{zz}, \tau_{zz}=g^\prime (h-\eta)$, and applying the Leibnitz rule to the second terms in the RHS of equations (\ref{x-mom}) and (\ref{y-mom}).

In the following section, we further develop Castro-Orgaz \emph{et al}.'s theory into a refined full non-hydrostatic shallow granular flow model.

\section{Further development of non-hydrostatic shallow granular flow theory}

We first reduce the vertical normal stress formula (\ref{tauzz_0}) to a polynomial form in the relative elevation $\eta$.  The result will show that Denlinger and Iverson's model \cite{Denlinger2004} is a special case of Castro-Orgaz \emph{et al.}'s theory \cite{Castro2014}. Then we calculate the normal stress acting on the bed according to the prescribed friction law. These calculations will lead to a refined full non-hydrostatic shallow granular flow model provided that the required  constitutive relations are prescribed. In the end we transform the full model into a form similar to Boussinesq water wave equations {presumedly more convenient} for numerical studies.

\subsection{Reformulation of vertical normal stress}\label{sec:reformu}

The role of $\tau_{zz}(\eta)$ in {equation} (\ref{tauzz_0}) is for evaluating $\bar{\tau}_{xx}$, $\bar{\tau}_{yy}$ and $\bar{\tau}_{xy}$, but this form is not convenient  for analytical integration in $\eta$, so we consider to simplify it.
The depth-averaged vertical velocity $\bar{w}$ is computed out from equation (\ref{wlinear2}) {for use in subsequent
derivation},
\begin{equation}
\label{wav2}
\bar{w}\equiv \frac{1}{h}\int_b^s w\mbox{d} \zeta=w_b-\left(\nabla \cdot \mathbf{\bar{u}}\right) \frac {h}{2}.
\end{equation}
Define $\hat{I}\equiv \int_b^z w\mbox{d} \zeta=\int^s_b w\mbox{d}\zeta -\int^s_z w \mbox{d} \zeta =h\bar{w}-I$, and
rewrite equation (\ref{tauzz_0}) as
\begin{eqnarray}
\label{tauzz2}
\tau_{zz}&=&  \rho g (h-{\color{blue}\eta}) + \rho h\left(\frac{\partial \bar{w} }{\partial t} +\mathbf{\bar{u}} \cdot \nabla \bar{w}\right)  -\rho \left[ \frac{\partial \hat{I}}{\partial t} +\nabla\cdot (\bar{\mathbf{u}} \hat{I})\right] - \rho w^2,
\end{eqnarray}
where the volume conservation equation  (\ref{depth2}) has been used. $\hat{I}$ is computed from  (\ref{wlinear2}) as
\begin{equation}
\label{Idown}
\hat{I}=w_b {\color{blue}\eta}-(\nabla \cdot \mathbf{\bar{u}} )\frac{\color{blue}\eta^2}{2}.
\end{equation}
Insert (\ref{Idown}) into  (\ref{tauzz2}), and denote the total time derivative  $D/Dt=\partial_t+  \mathbf{\bar{u}}\cdot \nabla $, we obtain
\begin{eqnarray}
\tau_{zz}&=&\rho g (h-{\color{blue}\eta}) + \rho h\frac{D\bar{w} }{Dt}  -\rho \left\{\underbrace{ \frac{\partial w_b}{\partial t}{\color{blue}\eta}}_{\text{to 1}}-\underbrace{w_b \frac{\partial b}{\partial t} }_{\text{to 2}} +(\nabla \cdot \mathbf{\bar{u}}) w_b{\color{blue}\eta}+\underbrace{\bar{u}\frac{\partial (w_b {\color{blue}\eta})}{\partial x}  +\bar{v}\frac{\partial (w_b{\color{blue}\eta})}{\partial y} }_{ \text{expand and to 1 and 2} }  \right. \nonumber \\
 &&\left.  -\underbrace{\frac{\partial (\nabla \cdot \bar{\mathbf{u}}) }{\partial t}\frac{\color{blue}\eta^2}{2}}_{\text{to~}3} +\underbrace{ (\nabla \cdot \mathbf{\bar{u}}){\color{blue}\eta}\frac{\partial b}{\partial t}}_{\text{to~4}}  -(\nabla \cdot \mathbf{\bar{u}})^2 \frac {\color{blue} \eta^2}{2}-\underbrace{\frac{\bar{u}}{2} \frac{\partial \left[(\nabla \cdot \mathbf{\bar{u}}) {\color{blue}\eta^2}\right]  }{\partial x}
 -\frac{\bar{v}}{2} \frac{\partial \left[(\nabla \cdot \mathbf{\bar{u}}) {\color{blue}\eta^2}\right] }{\partial y}}_{\text{expand and to~3 and 4}} \right\}-\rho w^2  \nonumber  \\
&=&\rho g (h-{\color{blue}\eta}) + \rho h \frac{D \bar{w} }{D t} -\rho \left[\underbrace{\frac{D w_b}{D t}{\color{blue}\eta}}_1 -\underbrace{w_b^2}_2  - \underbrace{\frac{D (\nabla \cdot \bar{\mathbf{u}})}{Dt}\frac{\color{blue}\eta^2}{2}}_3
+ \underbrace{ (\nabla \cdot \mathbf{\bar{u}}){\color{blue}\eta}w_b}_4 \right. \nonumber \\
 && \left. + (\nabla \cdot \mathbf{\bar{u}}) w_b{\color{blue}\eta} -(\nabla \cdot \mathbf{\bar{u}})^2 \frac {\color{blue} \eta^2}{2}\right]- \rho w^2  \nonumber  \\
&=&\rho g (h-{\color{blue}\eta}) +  \rho \frac{D\bar{w} }{Dt} (h\underbrace{-{\color{blue}\eta}  )}_{\text{from 1} }  -\rho \left[\underbrace{\frac{D (h \nabla \cdot \mathbf{\bar{u}})}{Dt}\frac{{\color{blue}\eta}}{2}}_{\text{from 1 by (\ref{wav2})}} \underbrace{ -w_b^2+ 2(\nabla \cdot \mathbf{\bar{u}}) w_b{\color{blue}\eta}-(\nabla \cdot \mathbf{\bar{u}})^2¡¡{\color{blue} \eta^2} }_{=-w^2 \text{~by (\ref{wlinear2}) }}     \right. \nonumber \\
 && \left.  -\frac{D (\nabla \cdot \bar{\mathbf{u}})}{Dt}\frac{\color{blue}\eta^2}{2}  +(\nabla \cdot \mathbf{\bar{u}})^2 \frac {\color{blue} \eta^2}{2}\right]- \rho w^2     \nonumber \\
&=& \underbrace{ \rho g (h-{\color{blue}\eta})}_\text{hydrostatic} +\underbrace{ \rho \frac{D \bar{w} }{D t} (h-{\color{blue}\eta}) }_\text{mean acceleration cor.} -\underbrace{\frac{\rho}{2}\left[\frac{D (h \nabla \cdot \mathbf{\bar{u}})}{Dt}{\color{blue}\eta}  - \frac{D (\nabla \cdot \bar{\mathbf{u}})}{Dt} {\color{blue}\eta^2 } + (\nabla \cdot \mathbf{\bar{u}})^2{\color{blue}\eta^2 } \right]}_\text{high order acceleration correction}.
\label{tauzz0-new}
\end{eqnarray}
It is seen that $\tau_{zz}$ equals to a hydrostatic pressure of the order of $\rho g H$ plus a {mean} vertical  acceleration correction term of the order of $\rho gH $ and a high order {acceleration} correction term of the order of $\rho g H \epsilon $. The last  term is parabolic in $\eta$ and becomes zero at both the basal surface $\eta=0$ and the free surface $\eta=h$.  Note that the first two terms are {the same linear distribution of $\tau_{zz}$ as}  in \cite{Denlinger2004}. Further, the depth-averaged vertical velocity (\ref{wav2}) is identical to the arithmetic average of vertical velocities between the basal and free surfaces, equation (\ref{wav}). Therefore, the first two terms are completely identical to Denlinger and Iverson's {$\tau_{zz}(z)$},  and their model, if written in a differential form, differs from Castro-Orgaz \emph{et al.}'s theory only in the last term in equation (\ref{tauzz0-new}).

\subsection{Basal traction vector calculation}

Noting that the outward unit vector normal to the bed is $\mathbf{n}_b=(\partial_x b, \partial_y b,-1) /\Delta_b$, where   $\Delta_b=[1+(\partial b/\partial x)^2 +(\partial b/\partial y)^2  ]^{1/2}$ is the normalisation factor, the basal traction vector $\mathbf{T} = (T_x, T_y, T_z)=\bm\tau_b \cdot \mathbf{n}_b$ can be written as
\begin{equation}
\bm\tau_b \cdot \mathbf{n}_b = \frac{1}{\Delta_b } \left(\begin{array}{c}
\displaystyle \tau_{xx} \frac{\partial b}{\partial x}+ \tau_{yx} \frac{\partial b}{\partial y} -\tau_{zx}   \\
\displaystyle \tau_{xy} \frac{\partial b}{\partial x}+ \tau_{yy} \frac{\partial b}{\partial y} -\tau_{zy}  \\
\displaystyle \tau_{xz} \frac{\partial b}{\partial x}+ \tau_{yz} \frac{\partial b}{\partial y} -\tau_{zz} \end{array} \right)_b.   \label{traction}
\end{equation}
As noted in Refs.~\cite{{Gray2003},{Mangey2003}},
the RHS terms in equations (\ref{x-mom2}) and (\ref{y-mom2}) are the horizontal components of the basal traction vector. The vertical component of the basal traction vector occurs in the integration of the $z$-component equation of (\ref{mom}) from $z=b$ to $z=s$ by using Leibnitz's rule and  boundary conditions (\ref{kine1}),(\ref{kine2}), and (\ref{dy1}),
\begin{eqnarray}
\left(\tau_{zz} -\tau_{xz}\frac{\partial b}{ \partial x} -\tau_{yz}\frac{\partial b}{\partial y}\right)_b &=&  \rho g h +\rho \left[ \frac{\partial }{\partial t}\int_b^s w \text{d} \zeta +  \frac{\partial }{\partial x}\int_b^s uw \text{d} \zeta  + \frac{\partial }{\partial y}\int_b^s vw \text{d}\zeta \right]  \nonumber    \\
&&+    \frac{\partial }{\partial x} \int_b^s \tau_{xz} \mbox{d} \zeta + \frac{\partial}{\partial y} \int_b^s \tau_{yz} \text{d} \zeta. \label{z-momfull}
\end{eqnarray}
With the assumption of constant profile for $u$ and $v$, equation (\ref{z-momfull}) becomes
\begin{eqnarray}\label{z-momfull2}
\left(\tau_{zz} -\tau_{xz}\frac{\partial b}{ \partial x} -\tau_{yz}\frac{\partial b}{\partial y}\right)_b &=&  \rho g h +\rho h\frac{D \bar{w}}{Dt} + \frac{\partial (h \bar{\tau}_{xz}) }{\partial x}+ \frac{\partial (h \bar{\tau}_{yz}) }{\partial y} .
\end{eqnarray}
If $\tau_{zz}|_b,\tau_{xz}|_b,\tau_{yz}|_b, \bar{\tau}_{xz}$ and $\bar{\tau}_{yz}$ are $\mathcal{O}(\rho g H)$, $ \bar{u},\bar{v}$, and $\bar{w}$ are $\mathcal{O}(\sqrt{gL})$, $t$ is $\mathcal{O}(\sqrt{L/g})$, and  $\partial b/\partial x $ and $\partial b/\partial y$ are $\mathcal{O}(1)$, then the two shear stress terms in the RHS in (\ref{z-momfull2}) are $\mathcal{O}(\rho g H \varepsilon)$, while all other terms are $\mathcal{O}(\rho g H)$. Therefore, the two shear stress terms in the RHS in (\ref{z-momfull2})  can  be neglected and the equation  becomes
\begin{equation}
\left(\tau_{zz} -\tau_{xz} \frac{\partial b}{ \partial x} -\tau_{yz}\frac{\partial b}{\partial y} \right)_b  =\rho h \left( g + \frac{D\bar{w}}{D t}\right) =\rho   g^\prime  h.
\label{stressb}
\end{equation}
Multiply  the Coulomb friction law (\ref{dy2}) with $\Delta_b$ and expand the three components in the $x$, $y$ and $z$ directions, respectively,
\begin{eqnarray}
\left(\tau_{xx}\frac{\partial b}{\partial x}+\tau_{yx}\frac{\partial b}{\partial y}-\tau_{zx}\right)_b&=&\left(\mathbf{n}_b\cdot  \bm{\tau}_b\cdot \mathbf{n}_b\right)   \left[\frac{\Delta_b u_r}{|\mathbf{v}_{r}| } \tan\phi_\text{bed} + \frac{\partial b}{\partial x}  \right], \label{x-fric} \\
\left(\tau_{xy}\frac{\partial b}{\partial x}+\tau_{yy}\frac{\partial b}{\partial y}-\tau_{zy}\right)_b&=&\left(\mathbf{n}_b\cdot  \bm{\tau}_b\cdot \mathbf{n}_b\right)  \left[\frac{\Delta_b v_r}{|\mathbf{v}_{r}| } \tan\phi_\text{bed}+ \frac{\partial b}{\partial y}  \right], \label{y-fric} \\
\left(\tau_{xz}\frac{\partial b}{\partial x}+\tau_{yz}\frac{\partial b}{\partial y}-\tau_{zz}\right)_b&=&\left(\mathbf{n}_b\cdot  \bm{\tau}_b\cdot \mathbf{n}_b\right)  \left[\frac{\Delta_b w_r}{|\mathbf{v}_{r}| } \tan\phi_\text{bed}  -1    \right], \label{T-z}
\end{eqnarray}
where $\mathbf{n}_b\cdot  \bm{\tau}_b\cdot \mathbf{n}_b=\mathbf{n}_b\cdot \mathbf{T} $ is the normal stress acting on  the basal surface in the outward normal direction, and $\mathbf{v}_r=(u_r,v_r,w_r)$ is the (tangential) velocity difference at the bed.   By combining Eq.~(\ref{T-z})  with Eq.~(\ref{stressb}), we obtain the bed  normal stress
\begin{equation}
 \mathbf{n}_b\cdot  \bm{\tau}_b\cdot \mathbf{n}_b =\frac{\rho g^\prime h}{\displaystyle 1- \frac{\Delta_b w_r}{|\mathbf{v}_{r}| } \tan\phi_\text{bed} }= \rho \beta  g^\prime h,
\label{bednormalstress}
\end{equation}
where
\begin{equation}
\label{beta}
\beta=\frac{1}{  \displaystyle 1- \frac{\Delta_b w_r}{|\mathbf{v}_{r}| } \tan\phi_\text{bed}  }.
\end{equation}
Consequently, the horizontal components of the basal traction vector in equations (\ref{x-fric}) and (\ref{y-fric}) are
\begin{equation}
\begin{split}
&\left(\tau_{xx}\frac{\partial b}{\partial x}+\tau_{yx}\frac{\partial b}{\partial y}-\tau_{zx}\right)_b=\beta \rho g^\prime h   \left(\frac{\Delta_b u_r}{|\mathbf{v}_{r}| } \tan\phi_\text{bed}+ \frac{\partial b}{\partial x}  \right),
 \\
&\left(\tau_{xy}\frac{\partial b}{\partial x}+\tau_{yy}\frac{\partial b}{\partial y}-\tau_{zy}\right)_b= \beta \rho g^\prime h \left(\frac{\Delta_b v_r}{|\mathbf{v}_{r}| } \tan\phi_\text{bed} + \frac{\partial b}{\partial y}  \right).
\end{split}
\label{x-fric2}
\end{equation}
We remark that the normal stress acting on the bed can also be calculated   directly from expansion of  $\mathbf{n}_b\cdot  \bm{\tau}_b\cdot \mathbf{n}_b$ rather than from equation (\ref{bednormalstress}) as long as a constitutive equation and a velocity profile in $z$ are given.  In cases when a basal friction law (Coulomb, Manning) is given,   the use of equation (\ref{bednormalstress}) for calculating the bed normal stress is natural and simpler.

\subsection{A refined full non-hydrostatic shallow granular flow model}

To close  equations (\ref{x-mom2}) and (\ref{y-mom2}), lateral normal and shear stresses, $\tau_{xx},\tau_{yy},\tau_{xy}$ and $\tau_{yx}$  have to be parameterized. Savage and Hutter {\cite{Savage1989} proposed to use the Mohr-Coulomb soil constitutive law for the avalanche materials in the shallow-water continuum model. The lateral shear stresses $\tau_{xy}$ and $\tau_{yx}$ are omitted, and the lateral normal stresses $\tau_{xx}$ and $\tau_{yy}$ are related to the normal stress  $\tau_{zz}$ in the depth direction in standard fashion through the use of earth pressure coefficients $k_x$ and $k_y$ respectively. However, the method of determining $k_x$ and $k_y$ \cite{{Savage1989},{Wieland1999},{Wang2004}} assumes that
two principal axes of the stress tensor are in the $x$ and $y$ directions, This \emph{ad hoc} assumption destroys the rotational invariance of the equations about the $z$ direction perpendicular to the tangential plane. To amend this deficit, a variety of models have been proposed \cite{{Detoni2005},{Kelfoun2005},{Luca2009}}. For example, Chen \emph{et al}. \cite{Chen2013}, based on Ref.~\cite{Detoni2005}, use a symmetric earth pressure coefficient matrix $\mathbf{K}$ that is diagonalizable by rotating the coordinates with an invertible rotation matrix $\mathbf{T} =\left(\begin{array} {cc} \cos \gamma  & -\sin \gamma   \\  \sin \gamma  &   \cos \gamma   \end{array}\right)  $ such that
\begin{equation*}
\mathbf{T}^{-1}\mathbf{K}\mathbf{T}=\left(\begin{array}{cc}
k_1 & 0  \\ 0  &  k_2 \end{array} \right), \text{~~}  \left(\begin{array}{cc} \tau_{xx} & \tau_{xy} \\
\tau_{yx}  &  \tau_{yy}\end{array} \right) = \tau_{zz} \mathbf{K} ,
\end{equation*}
where $\gamma$ is the angle between the primary principal axis (assume to be parallel to the local flow velocity \cite{Detoni2005})  and the $x$-axis, and $k_1$ and $k_2$ are the primary and secondary earth pressure coefficients depending on the basal and internal friction angles and on the dilation/compaction states in principal axis directions \cite{Savage1989}.
In this work}, we use the isotropic lateral normal stresses and  a relation between the lateral shear and normal stresses deduced from the Coulomb equation \cite{Iverson2001},
\begin{equation} \label{lateralstress} {
\begin{split}
&\tau_{xx}=\tau_{yy}=k_\text{ap}\tau_{zz}, k_\text{ap}=k_x=k_y =2\frac{1 \mp \sqrt{ 1- \cos^2\phi_\text{int}/\cos^2 \phi_\text{bed} } }{\cos^2 \phi_\text{int}}  -1  ,  \frac{\partial \bar{u}}{\partial x}+ \frac{\partial \bar{v} }{\partial y}  \gtrless 0 ,\\
&\tau_{xy}=\tau_{yx}=-\text{sgn}\left(\frac{\partial \bar{u}}{\partial y} + \frac{\partial \bar{v}}{\partial x}\right) k_\text{ap}  \tau_{zz}\sin\phi_\text{int}=S\tau_{zz}, \end{split}
}
\end{equation}
{where $S=- \text{sgn}\left(\frac{\partial \bar{u}}{\partial y} + \frac{\partial \bar{v}}{\partial x}\right) k_\text{ap} \sin\phi_\text{int}$. The second row in equation (\ref{lateralstress}) is slightly modified from Ref. \cite{Iverson2001} in order to ensure $\tau_{xy}=\tau_{yx}$. Equation (\ref{lateralstress}) is rotationally invariant with respect to the $z$-axis.  With the constitutive relations available,}  the depth-averaged lateral normal  and {shear} stresses $\bar{\tau}_{xx}$, $ \bar{\tau}_{yy}$, $\bar{\tau}_{xy}$ and $\bar{\tau}_{yx}$ in momentum equations (\ref{x-mom2}) and  (\ref{y-mom2}) only require integration of {$\tau_{zz}(\eta)$ from equation  (\ref{tauzz0-new})}, which can be done analytically after denoting $\Phi=\nabla \cdot \mathbf{\bar{u}}$. The basal-traction horizontal components in the RHS of equations (\ref{x-mom2}) and  (\ref{y-mom2}) are evaluated with equation (\ref{x-fric2}). With these terms available, we transform equations (\ref{depth2})-(\ref{y-mom2})  into refined full non-hydrostatic shallow granular flow equations
\begin{equation}
\begin{split}
\frac{\partial}{\partial t}\left[\begin{array}{c} h  \\  h\bar{u}   \\  h\bar{v}  \end{array}\right] &+ \frac{\partial}{\partial x} \left[\begin{array}{c} h\bar{u} \\ h\bar{u}^2 + k_x \left[\displaystyle\frac12 g^\prime h^2 + \frac {h^3}{12}\left( \Phi^2 -\frac{D \Phi}{Dt}   \right)  \right] \\  h\bar{u}\bar{v} +\displaystyle S  \left[\frac 12 g^\prime h^2 +\displaystyle \frac {h^3}{12}\left( \Phi^2 -\frac{D \Phi}{Dt}  \right) \right] \end{array} \right]
\\
&
+\frac{\partial }{\partial y} \left[\begin{array}{c} h\bar{v} \\ h\bar{u}\bar{v} +\displaystyle S\left[\frac 12 g^\prime h^2 +\displaystyle \frac {h^3}{12}\left( \Phi^2 -\frac{D \Phi}{Dt}  \right) \right]   \\  h\bar{v}^2 + k_y\displaystyle\left[ \frac 12 g^\prime h^2 +\displaystyle \frac {h^3}{12}\left( \Phi^2 -\frac{D \Phi}{Dt}  \right)  \right]  \end{array} \right]  \\
&= \left[ \begin{array}{c}   0 \\
\displaystyle - \beta \rho g^\prime h   \left(\frac{\Delta_b u_r}{|\mathbf{v}_{r}| } \tan\phi_\text{bed}+ \frac{\partial b}{\partial x}  \right)  \\
\displaystyle - \beta \rho g^\prime h \left(\frac{\Delta_b v_r}{|\mathbf{v}_{r}| } \tan\phi_\text{bed} + \frac{\partial b}{\partial y}  \right) \end{array} \right],
\end{split}
\label{refined}
\end{equation}
where the $h^3$ terms in the depth-averaged vertical normal stress has been simplified by using the volume conservation equation
(\ref{depth2}).  System (\ref{refined}) can be further cast into a frequently used form of Boussinesq-type water wave equations (e.g., \cite{{Kim2009},{Fang2014}}) as follows. If we absorb the time partial derivatives in the convective fluxes into $\partial_t (hu,hv)$  by making use of equation (\ref{depth2}) and the assumption that $k_x,k_y$ and $S$ can be extracted out of the differential operators,  we can obtain (overbars in $\bar{u},\bar{v}$ and $\bar{w}$ have been omitted in the following context to simplify notations)
\begin{equation}
\label{waterwaveeq}
\frac{\partial \mathbf{U}}{\partial t} +
\frac{\partial \mathbf{F}}{\partial x} +
\frac{\partial \mathbf{G}}{\partial y}=\mathbf{S}_{f-b}+\mathbf{S}_d,
\end{equation}
where
\begin{eqnarray}
&&\mathbf{U}=\left[\begin{array}{c}   h  \\ ¡¡U   \\ V \end{array}  \right]=\left[\begin{array}{c}   h  \\ \displaystyle
hu+{\frac {k_x}{2}\frac{ \partial \left(h^2w\right)}{\partial x}+\frac {S}{2}\frac{\partial \left(h^2w\right)}{\partial y}} -\frac{k_x}{12}\frac{\partial\left(h^3\Phi\right)}{\partial x} -\frac{S}{12} \frac{\partial \left(h^3\Phi\right)}{\partial y} \\ \displaystyle
hv+{\frac {S}{2} \frac{\partial\left(h^2w\right)}{\partial x}+\frac {k_y}{2}\frac{ \partial\left(h^2w\right)}{\partial y}}  -\frac{S}{12} \frac{ \partial\left(h^3\Phi\right)}{\partial x}- \frac{k_y}{12}\frac{ \partial\left(h^3\Phi\right)}{\partial y} \end{array}
  \right] , \nonumber  \\
&&   \label{newU}              \\
&&\mathbf{F}=\left[\begin{array}{c}  hu \\ \displaystyle  hu^2+ \frac 12   gh^2  \\  \displaystyle huv  \end{array} \right],~~
\mathbf{G}=\left[\begin{array}{c}  hv  \\ \displaystyle  huv \\  \displaystyle   hv^2  + \frac 12   gh^2   \end{array} \right],
\end{eqnarray}
\begin{eqnarray}
&& \mathbf{S}_{f-b}= \left[ \begin{array}{c}   0 \\
\displaystyle - \beta \rho g^\prime h   \left(\frac{\Delta_b u_r}{|\mathbf{v}_{r}| } \tan\phi_\text{bed}+ \frac{\partial b}{\partial x}  \right)  \\
\displaystyle - \beta \rho g^\prime h \left(\frac{\Delta_b v_r}{|\mathbf{v}_{r}| } \tan\phi_\text{bed} + \frac{\partial b}{\partial y}  \right) \end{array} \right], ~~  \mathbf{S}_{d}=-\left[ \begin{array}{c}   0 \\
\displaystyle  k_x \frac{\partial \Gamma}{\partial x} + S\frac{\partial \Gamma}{\partial y}   \\
\displaystyle  S\frac{\partial \Gamma}{\partial x}+k_y\frac{\partial \Gamma}{\partial y} \end{array} \right] ,
\\
&&\Gamma={\frac {1}{2} \left[  u\frac{\partial(h^2w)}{\partial x}+  v\frac{\partial(h^2w)}{\partial y}+2h^2w\Phi\right] }  -\frac{1}{12}\left[u\frac{\partial\left(h^3\Phi\right)}{\partial x}+v\frac{\partial \left(h^3\Phi\right)}{\partial y} +2h^3\Phi^2\right] \nonumber. \\
\end{eqnarray}
Here, $\mathbf{S}_{f-b}$ are the  friction and  bed slope terms, and $\mathbf{S}_d$ is the dispersive terms. However, we can see
the lumped conservative variable vector $\mathbf{U}$ and the dispersive terms are still  complicated,  so we leave numerical solution of equations (\ref{waterwaveeq}) to  future research. In the remaining parts of this paper, we will focus on numerical solution of a simplified model deduced from system (\ref{refined}) using an software.

\section{A simplified  non-hydrostatic shallow granular flow model}

Our initial simplified non-hydrostatic shallow granular flow model results from system (\ref{refined}) by neglecting all the $h^3$ terms in the convective fluxes, which is the same as the model \cite{Denlinger2004} except  slightly different
lateral normal-shear stress relation (\ref{lateralstress}) and basal normal stress (\ref{bednormalstress}).
However, we encountered numerical instability problem when implementing this initial model (also the model \cite{Denlinger2004}) on TITAN2D \cite{titancode}. Therefore, we try to find an approximate formula for the enhanced gravity to be given in Sec.~\ref{sec:enhanced}, and correspondingly, we add a  ``centripetal
normal stress" due to the curvature tensor to the original basal normal stress in the RHS terms. Our final simplified model is presented in  Sec.~\ref{sec:reduced model}.

\subsection{Enhanced gravity}\label{sec:enhanced}

In implementing the Denlinger and Iverson model \cite{Denlinger2004} on TITAN2D, we experienced that the enhanced gravity $g^\prime$  (\ref{acce}) posed difficulty for numerical solution. Specifically, when evaluating $g^\prime$,  the finite difference approximation for $\partial {\bar{w}}/{\partial t}$
often causes numerical instability or irregularity. Therefore, {we derive} an approximate formula for $g^\prime$ by letting the bed normal stress (\ref{bednormalstress}) equal to the traditional hydrostatic bed normal stress obtained from the shallow flow argument in a bed-fitted coordinate system, {as described below}.

\begin{figure}[h]
\centering
\includegraphics[width=16 pc  ]{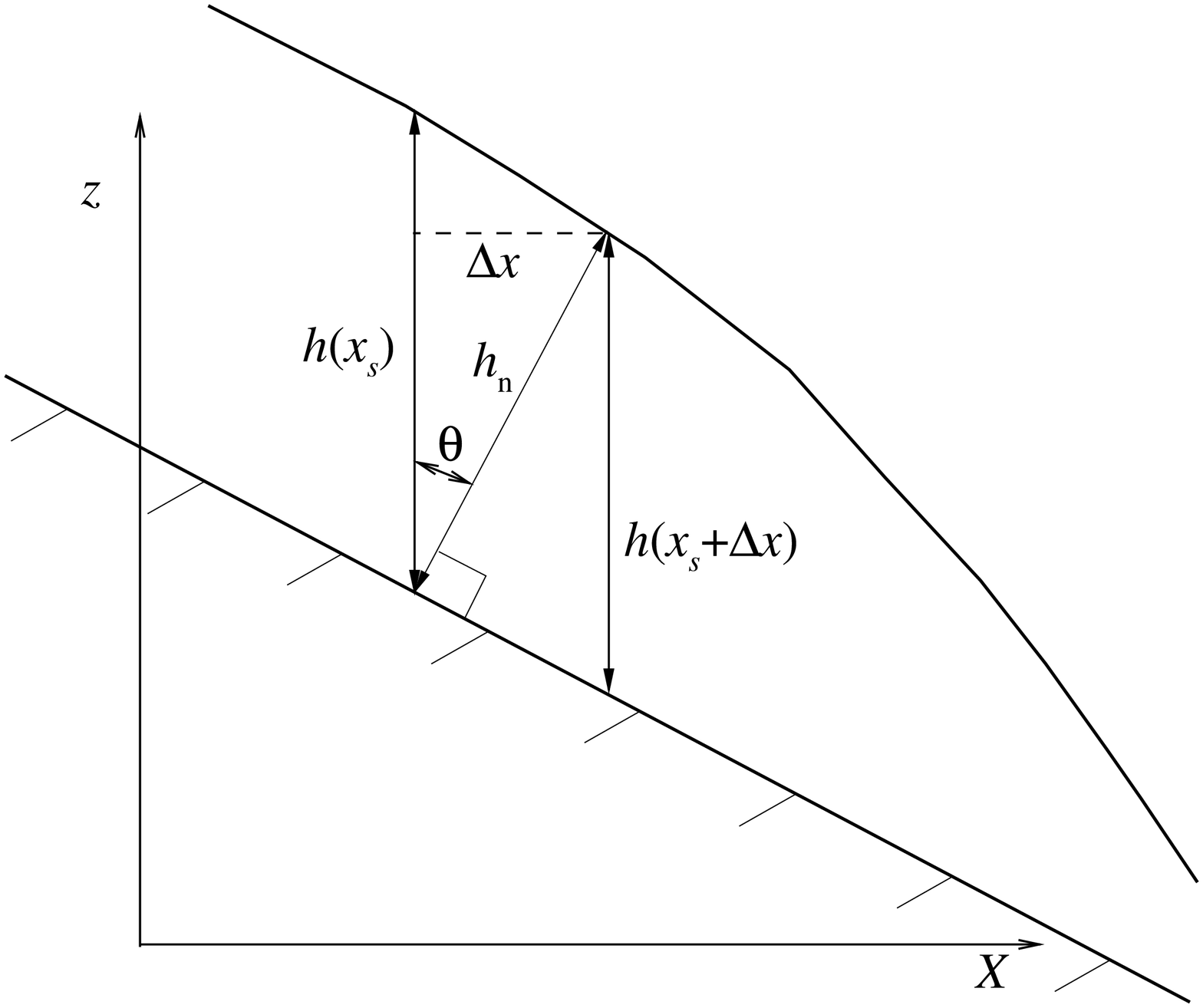}
\caption{Relation between vertical and normal depths, where $\Delta x= h_\text{n} \sin \theta$, and $X$ is in the horizontally  projected  steepest downslope direction.  }
\label{figure2}
\end{figure}

Based on scaling analysis of equations written in a local Cartesian coordinate system with the $\tilde{z}$ axis normal to the bed \cite{{Savage1989},{Gray2003},{Iverson2001},{Pitman2005}}, the bed normal stress balances the normal component
of the mass weight if neglecting curvature effects,
\begin{equation}
\mathbf{n}_b\cdot  \bm{\tau}_b\cdot \mathbf{n}_b =\rho g h_\text{n} \cos\theta. \label{force-choice3}
\end{equation}
Here, $\theta$ is the angle between the vertical $z$-axis and the normal to the bed, and $h_\text{n}$ is the depth in the bed normal direction, see figure \ref{figure2}. If the basal surface is regarded as a planar surface in proximity of position $x_s$, then there is a geometrical relation between  the vertical depth $h(x_s+\Delta x) $ at $x_s+\Delta x$ and the normal depth $h_\text{n}$ at $x_s$,
\begin{equation}
 h_\text{n}= h(x_s+\Delta x) \cos \theta
\end{equation}
where $\Delta x =h_\text{n} \sin \theta $.
Using the first order Taylor expansion with respect to position $x_s$, we obtain
\begin{equation*}
h_\text{n}  \approx \left(\displaystyle  h (x_s) + h_\text{n} \sin \theta \left.\frac{\partial h}{\partial X}\right|_{x_s} \right) \cos \theta ,
\end{equation*}
i.e.,
\begin{equation}
 h_\text{n}=  \frac{h\cos \theta}{1-\displaystyle \left.\frac{\partial h}{\partial X}\right|_{x_s}\tan\theta \cos^2 \theta },
\label{hrela}
\end{equation}
where $\partial /\partial X=-\left(\partial_x b/\sqrt{(\partial_x b)^2+(\partial_y b)^2}\right)\partial  /\partial x - \left(\partial_y b/\sqrt{(\partial_x b)^2+(\partial_y b)^2}\right)\partial  /\partial y$ is the directional derivative in the horizontal plane in the steepest downslope direction. Note that equation (\ref{hrela}) takes account for variation of $h$ in space, thus is expected to be more accurate than $h_\text{n}=h\cos \theta $ valid for uniform depth as given by
Juez \emph{et al}. \cite{Juez2013}.

Now, let equation (\ref{bednormalstress}) equal to  equation (\ref{force-choice3}) together with (\ref{hrela}). Noting that $\tan\theta=\sqrt{(\partial_x b)^2+(\partial_y b)^2}$, we obtain an {approximate formula for the enhanced gravity}
\begin{equation}
\label{gravity-A}
g^\prime= \frac{g \left(\displaystyle 1- \frac{\Delta_b w_r}{|\mathbf{v}_{r}| } \tan\phi_\text{bed}\right )\cos^2\theta } {\displaystyle  1+\left( \frac{\partial h}{\partial x} \frac{\partial b}{\partial x} +
\frac{\partial h}{\partial  y} \frac{\partial b}{\partial y} \right) \cos^2\theta},
\end{equation}
which does not involve time derivative occurred in equation (\ref{acce}). We find this $g^\prime$ is numerically more stable than equation (\ref{acce}).   The only disadvantage lies in the fact that $\partial h/ \partial x$ and $ \partial h/ \partial y$ might be very large near shock waves or initial jumps, making the denominator approach zero. In our computation, to avoid this  problem, a varying under-relaxation factor, $ \exp (-\omega |\mathbf{v}_r|| \nabla h |$), is multiplied in front of the second term in the  denominator of equation (\ref{gravity-A}), where $\omega$ is a free parameter (tuned in $1\sim 3$ in this work). It is seen that when the magnitude of basal velocity difference  $\mathbf{v}_r$ or gradient $\nabla h$  is large  (e.g., supercritical flows with shocks or large initial jumps), this factor becomes small so as to suppress destabilization. On the other hand,  when $\nabla h$ is small or $\mathbf{v}_r$ approaches 0, this factor becomes 1 so as to  recover the original equation (\ref{gravity-A}).

The rationale for bed normal stress (\ref{bednormalstress}) together with  {enhanced gravity}   (\ref{gravity-A}) can be verified by a simple one-dimensional granular flow example of uniform thickness descending a frictionless slope inclined
at a constant angle $\theta$ \cite{Denlinger2004}. Since $\phi_\text{bed}=0$ and $\partial h /\partial x=0$, equation (\ref{gravity-A}) gives  $g^\prime=g\cos^2\theta$ which is the same as that in \cite{Denlinger2004},  and equation (\ref{bednormalstress}) gives $\mathbf{n}_b\cdot  \bm{\tau}_b\cdot \mathbf{n}_b=\rho g h\cos^2\theta=\rho g h_\text{n} \cos \theta$ which is evidently correct.  Another example to verify the correctness of equations (\ref{bednormalstress}) and  (\ref{gravity-A}) is the 1D static steady state of shallow water flows. For such a state, $h+b=\text{const}$, hence $\partial h/\partial x=- \partial b/\partial x =\tan \theta $, and equation (\ref{gravity-A}) gives $g^\prime=g$ so that equation (\ref{bednormalstress}) recovers the traditional hydrostatic basal pressure $\rho g h$. The present approximate $h_\text{n}$ (\ref{hrela}) along with equation (\ref{force-choice3}) also {recovers  $\rho g h$}.  On the other hand, the approximation of $h_\text{n}=h\cos \theta $ as in Ref. \cite{Juez2013} together with equation (\ref{force-choice3}) will lead to a basal pressure of $\rho g h \cos^2\theta$, which is incorrect {for this state}.

\subsection{Governing equations of the simplified non-hydrostatic model}\label{sec:reduced model}

We restrict ourself to a fixed bed (i.e., $\partial b/\partial t=0$) in the following context. For the basal friction terms in the RHS of system (\ref{refined}),  the basal velocity difference in the basal sliding friction law is $u_r\approx \bar{u}$, $v_r\approx \bar{v}$, and $w_r =w_b = \bar{u}b_x+\bar{v}b_y$, so that  $ |\mathbf{v}_r|=\sqrt{u_r^2+v_r^2+w_r^2}=\sqrt{\bar{u}^2+\bar{v}^2+(\bar{u}b_x+\bar{v}b_y)^2}$. {It is noted that the approximate enhanced gravity} (\ref{gravity-A})  does not reflect the curvature effects of terrains. One important effect of the curvature is to produce an additional friction force linked to
centrifugal acceleration. Following Refs. \cite{Bouchut2004} and \cite{{Mangey2007}}, we account for the curvature effects by adding a {centripetal force} term involving the curvature tensor $\mathcal{H}$ of the
bed profile, $ (\mathbf{u}^T\mathcal{H}\mathbf{u})h_\text{n}/c^2$, to the basal normal stress ($\beta g^\prime  h$) occurring in the RHS of the momentum equations. Here, $c=\cos\theta$, $h_\text{n}$ is the avalanche thickness in the bed normal direction estimated with equation (\ref{hrela}), and the curvature tensor \cite{Mangey2007} is
\begin{equation}
\mathcal{H}=c^3\left[\begin{array}{cc}\displaystyle  \frac{\partial^2 b}{\partial x^2 }   & \displaystyle   \frac{\partial^2 b}{\partial  x \partial y }  \\
\displaystyle  \frac{\partial^2 b}{\partial  x \partial y }  & \displaystyle  \frac{\partial^2 b}{\partial  y^2 } \end{array} \right].
\end{equation}
The final set of governing equations of the simplified non-hydrostatic model result from system (\ref{refined}) by dropping all $h^3$ terms in the LHS and adding the {centrifugal force due to curvature only to the basal normal stress} in the RHS.  The equations can be written
in vector  form (overbars in $u,v$ have been omitted for brevity)
\begin{equation}
\frac{\partial \mathbf{U}}{\partial t}+\frac{\partial \mathbf{F}}{\partial x} + \frac{\partial \mathbf{G}}{\partial y} =\mathbf{S}(\mathbf{U}), \label{conslaw-U}
\end{equation}
where
\begin{equation}
\begin{split}
&\mathbf{U}=\left(\begin{array}{c}  h \\ hu \\ hv \end{array} \right),~~  \mathbf{F}=\left(\begin{array}{c}
hu \\   hu^2 +\displaystyle \frac 12 k_x g^\prime h^2   \\  huv \end{array}\right),  \\
&\mathbf{G}=\left(\begin{array}{c}
hv \\   huv     \\  hv^2 +\displaystyle \frac 12 k_y g^\prime h^2 \end{array}\right), ~~ \mathbf{S}(\mathbf{U})=\left(\begin{array}{c}  0 \\   s_x \\  s_y \end{array} \right),
\end{split}
\label{uvector}
\end{equation}
and
\begin{equation}
\label{svector}
\begin{split}
s_x&= -\frac{S}{2}\frac {\partial (g^\prime h^2 )}{\partial y} -\left(\beta g^\prime h +\frac{\mathbf{u}^T\mathcal{H}\mathbf{u}}{c^2}h_\text{n}  \right)_{+} \left( \frac{u}{|\mathbf{v}_r| }\Delta_b \tan\phi_\text{bed} + \frac{\partial b}{\partial x}\right) ,   \\
s_y &= -\frac{S}{2}\frac {\partial (g^\prime h^2 )}{\partial x} -\left(\beta g^\prime h +\frac{\mathbf{u}^T\mathcal{H}\mathbf{u}}{c^2}h_\text{n}  \right)_{+} \left(\frac{v}{|\mathbf{v}_r| }\Delta_b \tan\phi_\text{bed}+\frac{\partial b}{\partial y} \right)  .
\end{split}
\end{equation}
The subscript ``+" stands for the positive part, $x_+ = \max(0, x)$, and $\mathbf{u}=(u,v)^T$. Note that the depth-averaged lateral shear stress terms have been placed to the first terms in the RHS of equation (\ref{svector}). They could be retained in convective fluxes  $\mathbf{F}$ and $\mathbf{G}$,  but we follow Refs. \cite{{Iverson2001},{Patra2005}} to attribute them to the RHS friction source terms. {This arrangement makes equations (\ref{conslaw-U}) look like the traditional shallow water equations in the  horizontal-vertical Cartesian coordinate system except a variable $g^\prime$}.

\section{Numerical Method}

\subsection{Finite Volume Method}

The governing equations (\ref{conslaw-U}) are solved with a  Godunov type
finite volume method  for solving hyperbolic conservation laws.
We use a horizontal Cartesian mesh to discretize the computational domain. The flow solution variables $(h, hu, hv)$ are cell averages on each rectangular mesh cell.
The finite volume method used is a second-order predictor-corrector Godunov method \cite{Davis1998} with van Leer MUSCL reconstruction for  $\mathbf{U}$. The
intercell numerical flux is computed with the HLL flux. The wet/dry front is treated by using the Riemann invariant of the wave emanating from the front \cite{Denlinger2001}. The source term $\mathbf{S}$ is evaluated in a pointwise way using cell average values.   The predictor-corrector scheme used in  open-source code TITAN2D is listed as follows.

Equation (\ref{conslaw-U})  can be rewritten as
\begin{equation}
\label{quasilinear}
\mathbf{U}_t+\mathbf{A}\cdot \partial_x \mathbf{U} + \mathbf{B} \cdot \partial_y \mathbf{U}=\mathbf{S}(\mathbf{U}),
\end{equation}
where $\mathbf{A}=\left(\partial \mathbf{F} / \partial \mathbf{U}\right)_{g^\prime=\text{Const}}$ and $\mathbf{B}=\left(\partial \textbf{G}/ \partial \textbf{U}\right)_{g^\prime=\text{Const}}$ are approximate Jacobian matrices of fluxes evaluated {with $g^\prime$ frozen at the previous time level},  which have familiar forms as in the literature (e.g., \cite{{Denlinger2001},{Wang2004}}).

Given $\mathbf{U}^{n}_{i,j}$, the $(i,j)$ cell average at time level $n$, the
middle time predictor step is:
\begin{equation}
\label{pred}
\mathbf{U}^{n+\frac12}_{i,j}=\mathbf{U}^{n}_{i,j}-\frac{\Delta t}{2} \left(\mathbf{A}^{n}_{i,j}{\Delta}_x
\mathbf{U}^n_{i,j}+\mathbf{B}^{n}_{i,j} {\Delta}_y \mathbf{U}^n_{i,j}- \mathbf{S}_{i,j}^n \right),
\end{equation}
where $\Delta t$ is the time step,  $\Delta_x \mathbf{U}$  and $\Delta_y \mathbf{U}$ are limited slopes of  $\mathbf{U}$ in the $x$ and $y$ directions,
respectively. The depth-averaged lateral shear stress terms in $\textbf{S}$ are expanded using the chain rule, e.g., $\partial (g^\prime h^2)/\partial y = h^2 \partial_y g^\prime +2 g^\prime h \Delta_y h $, and  the two partial derivatives are discretized  like $\Delta_y \mathbf{U}$.

In the corrector step, a conservation update of $\mathbf{U}$ is
computed as follows:
\begin{equation}
\mathbf{U}^{n+1}_{i,j}= \mathbf{U}^n_{i,j}-\frac{\Delta t}{\Delta x} \left(\mathbf{F}^{n+\frac 12}_{i+\frac 12}- \mathbf{F}^{n+\frac 12}_{i-\frac 12}\right) -\frac{\Delta t}{\Delta y} \left[\mathbf{G}^{n+\frac 12}_{j+\frac 12}- \mathbf{G}^{n+\frac 12}_{j-\frac 12}\right]
+\Delta t \mathbf{S}^{n+\frac 12}_{i,j} ,\label{corr}
\end{equation}
where $\mathbf{F}_{i+1/2}^{n+1/2}=\mathbf{F}^\text{HLL}(\mathbf{U}^l_{i+1/2},\mathbf{U}^r_{i+1/2})$, and the
left and right state values are obtained by the MUSCL  reconstruction of the cell average values to  the edge position; that is, $\mathbf{U}^l_{i+1/2}=\mathbf{U}^{n+1/2}_{i,j}+(\Delta x/2)\Delta_x \mathbf{U}^{n+1/2}_{i,j}$, and
$\mathbf{U}^r_{i+1/2}=\mathbf{U}^{n+1/2}_{i+1,j}-(\Delta x/2) \Delta_x \mathbf{U}^{n+1/2}_{i+1,j}$. Notice that the  mechanical behavior of a Coulomb material has to be taken into account when evaluating the basal friction force in $\mathbf{S}$.  The frictional force will be treated by a special procedure to be given in section \ref{sec:stopping cri}.

The above predictor-corrector scheme is implemented in  TITAN2D, which has been incorporated with parallel adaptive Cartesian meshes and geographic information
system (GIS) databases \cite{{Patra2005},titancode}.

\subsection{Admissible Friction and Stopping Criteria}\label{sec:stopping cri}

The granular material can remain static even with an
inclined free surface. This equilibrium is not automatically preserved by the
finite volume scheme   and a  special procedure has to be introduced in the numerical solution for
the particular case  when the magnitude of an admissible  tangential stress  vector $\mathbf{T}_t$ (or \emph{residual inertia}), is smaller than the Coulomb friction threshold $\tau_{\max}=  \beta g^\prime h \tan \phi_\text{bed} $. In the following,
we describe how to calculate the admissible tangential stress vector $\mathbf{T}_t$.
The procedure is similar to that in Ref. \cite{Mangey2003}. We take the corrector step (\ref{corr}) as an example. A similar procedure also applies to the predictor step (\ref{pred}).

The mass and momentum components in equation (\ref{corr}) for any mesh cell $i$ are
\begin{equation}
\label{mass-compo}
\begin{split}
&h_{i}^{n+1} = h^n_{i}+\mathcal{F}^{n+\frac 12}_{hi}, \\
&\mathbf{q}^{n+1}_{i}=\mathbf{q}^n_{i}+\mathcal{F}^{n+\frac 12}_{\mathbf{q}i}-\Delta t\mathbf{s}_{bi}^{n+\frac 12 } +\Delta t\mathbf{f}^{n+\frac 12 }_{i},
\end{split}
\end{equation}
where $\mathbf{q}=(hu,hv)$,  $\mathcal{F}$ is the flux difference terms, $\mathbf{s}_{b}$ is the bed slope source term, and $\mathbf{f}^{n+1/2}_i$ is the sum of lateral shear and basal friction terms in equation (\ref{svector}), which is to be quantified in the following special procedure.  Define
\begin{equation}
\tilde{\mathbf{q}}^{n+1}_{i}=\mathbf{q}^n_{i}+\mathcal{F}^{n+1/2}_{\mathbf{q}i}-\Delta t\mathbf{s}_{bi}^{n+\frac 12 },
\end{equation}
which is an intermediate  solution  without any friction term. $\tilde{\mathbf{q}}^{n+1}_i/\Delta t$ is the
horizontal components of the so-called driving force \cite{Mangey2003} or the admissible basal shear stress vector \cite{Mae2013} which is in the basal tangential direction, $\mathbf{T}_{ti}^{n+1}$. The magnitude of  $\mathbf{T}_{ti}^{n+1}$ is calculated based on horizontal components $\tilde{\mathbf{q}}^{n+1}_{i}=(\tilde{q}_{x,i}^{n+1}, \tilde{q}_{y,i}^{n+1})$,
\begin{equation}
\left | \mathbf{T}^{n+1}_{ti}\right| =\frac{1}{\Delta t}\sqrt{\left|\tilde{\mathbf{q}}_i^{n+1}\right |^2  +\left(\tilde{q}_{x,i}^{n+1} \frac{\partial b}{\partial x} +\tilde{q}_{y,i}^{n+1}\frac{\partial b}{\partial y}\right)^2}.
\end{equation}
The special procedure is as follows.
\begin{enumerate}
\item [ 1)]   If the magnitude  of the driving force  $\mathbf{T}_t$ is less than the Coulomb threshold $\tau_{\max}$,
    and the slope angle of the free surface is less than the internal friction angle, i.e.,
\begin{equation}
 \left|\mathbf{T}_{ti}^{n+1}\right| < \beta g^\prime h^{n+1/2}_i\tan\phi_\text{bed} ,\text{~and\quad}  |\nabla (h^{n+1/2}_i+b)  | <\tan \phi_\text{int},
 \end{equation}
 then  the mass stops, i.e., $\mathbf{q}_{i}^{n+1}=0$. Actual values of $\mathbf{f}^{n+1/2}_i$  are not needed.
\item [2)] Otherwise,  the total friction force $\mathbf{f}^{n+1/2}_i$ is computed using dynamic values and the solution $\mathbf{q}_{i}^{n+1}$  is updated by equation (\ref{corr}).
\end{enumerate}
The dynamic quantity  $(u,v)_i^{n+1/2} /  | \mathbf{v}_{ri}^{n+1/2}  | $ in the basal friction terms are replaced by   $(\tilde{q}_{x},\tilde{q}_{y})_i^{n+1} / | \tilde{\mathbf{q}}_i^{n+1} | $  only  when $\mathbf{u}^{n+1/2}_i= 0$   to avoid division by zero \cite{Mangey2003}.

\section{Numerical Examples}\label{sec:vv}

The present simplified model is implemented in TITAN2D code and tested in {a} dam break problem {having an} analytical solution, {an} avalanche problem over simple topography, and {a} granular avalanche problem in {the} laboratory. For convenience of discussion, we refer to model A as governing equations (\ref{conslaw-U}) with $g^\prime$ being (\ref{gravity-A}), model B as the same governing equations with $g^\prime$ being (\ref{acce}) but $\partial \bar{w}/\partial t $ is set zero  to make {the model run stably}, and  model C as the same  governing equations but with $g^\prime = 9.8\text{~m}/\text{s}^2$, $\beta=\Delta_b =1$, $(u,v) / |\mathbf{v}_r  | \to (u,v) /|\mathbf{u}| $.  All the models use $k_x=k_y=1$ except stated explicitly in figure \ref{figwieland}(e).  Model C is the conventional Saint-Venant equations in the horizontal Cartesian coordinate system except having additional lateral shear stress terms as  in equation (\ref{svector}).

\subsection {Analytical solution of dam break problem}

Mangeney \emph{et al}. \cite{Mangey2000} gave the analytical solution for a one-dimensional dam {break} problem over an inclined plane in terms of thickness $\tilde{h}$ normal to the bed and coordinate $\tilde{x}$ tangential to the bed.
Juez \emph{et al}. \cite{Juez2013}
obtained the analytical solution for a similar dam break problem but in terms of vertical thickness $h$ and horizontal coordinate $x$. We compare our calculations with the analytical solution \cite{Juez2013}  since the solutions are expressed in the same horizontal coordinate system.

The considered problem is an inclined plane, on which a granular mass of a constant thickness and infinite length in the positive $x$ direction is released from rest.
Let  $\theta$ be the constant slope angle ($\theta>\phi_\text{bed}$)  and $u$ the horizontal velocity. For 1D flows over a planar slope, $\partial_x b=\tan\theta$. Assuming $g^\prime$ is constant and $\phi_\text{int}=0, \beta=1$,  equation (\ref{conslaw-U}) reduces to
\begin{equation}
\label{nonconserve}
\begin{split}
 \frac{\partial h}{\partial t}+u \frac{\partial h}{\partial x}  + h\frac{\partial u}{\partial x} &  =0   , \\
 \frac{\partial u}{\partial t}+u \frac{\partial u}{\partial x} + g^\prime \frac{\partial h}{\partial x} &  = -g^\prime (\tan\theta-\tan\phi_\text{bed}).
\end{split}
\end{equation}
Juez \emph{et al}.
\cite{Juez2013} took $g^\prime=g\cos^2\theta$ to obtain the analytical solution. Denoting  $m=-g^\prime(\tan\theta -\tan \phi_\text{bed})$ and using the following change of variables \cite{Mangey2000},
\begin{equation}
\label{CHI}
\begin{split}
& \chi  = x-\frac 12 m  t^2,\quad  \tau=t,   \\
&\mathcal{U}  =u- mt  , \quad \mathcal{H}=h,
\end{split}
\end{equation}
equation  (\ref{nonconserve}) can be transformed into a homogeneous system of equations for a frictionless, horizontal dam break problem with gravity acceleration of $g^\prime$, of which the classic Ritter solution \cite{Ritter1892} gives  \begin{equation}\label{exactsolu}
  (h,u ) = \left\{
      \begin{array}{ll}
       ( h_0,  mt ), &   \chi >   c_0t   \\
      \displaystyle \left(  \frac{h_0}{9}\left(2+\frac{\chi}{c_0 t}\right)^2, \frac 23\left(\frac {\chi}{t}-c_0\right) +mt \right), &        -2c_0 t \leq   \chi  \leq   c_0  t  \\
        (0, \text{arbitrary value} ) ,  &  \chi  <-   2c_0 t \\
      \end{array}.
    \right.
\end{equation}
where $c_0=\sqrt{g^\prime  h_0}$, and $h_0$ is the initial upstream vertical thickness.

\begin{figure}[h!]
 \centering
 \noindent\includegraphics[width=6cm ]{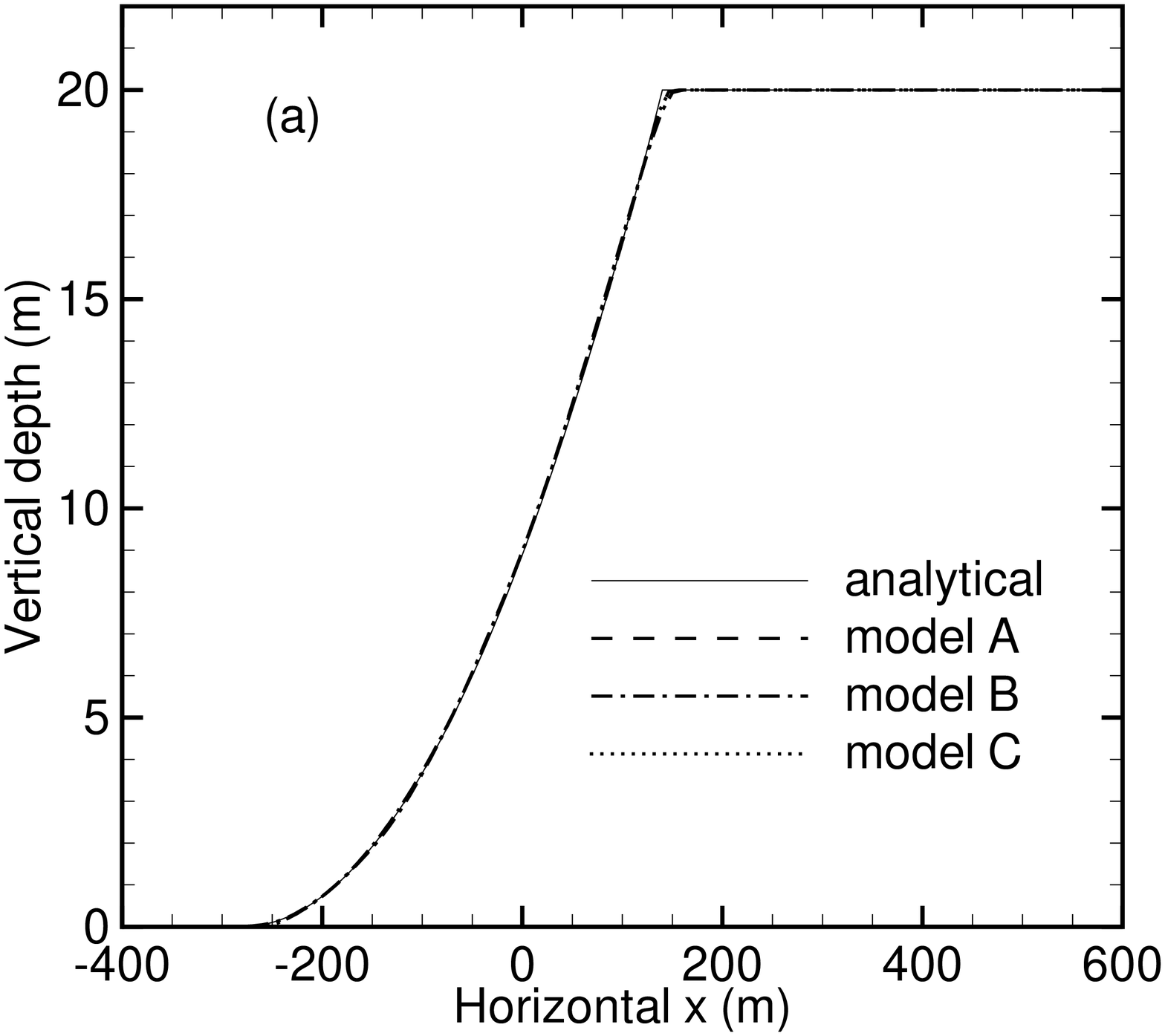}
 \noindent\includegraphics[width=6cm]{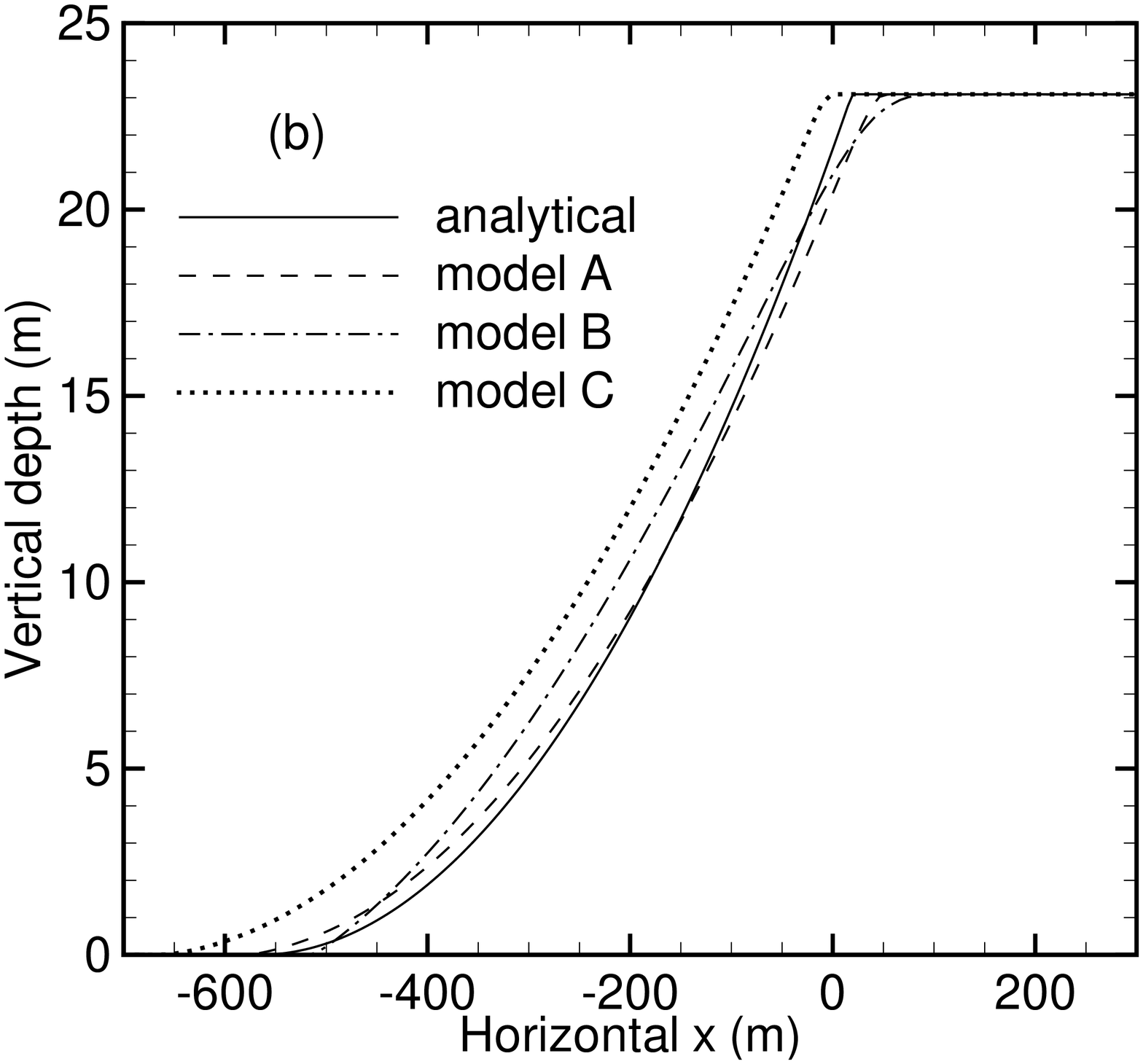}
\caption{Comparison of computed flow depth in the vertical direction versus horizontal distance from the initial edge of the dam at $x=0$  with the analytical solution. (a) Results for a tabular reservoir of sand with zero bed slope, zero internal friction,
and zero bed friction at $t=10$~s. (b) Results for a
tabular reservoir of sand with a 30$^\text{o}$ bed slope, zero internal
friction, and 20$^\text{o}$ basal friction at $t=15$~s. Meanings of models A, B and C are explained in the beginning  of section \ref{sec:vv}.
}
\label{dambreak}
\end{figure}

We compare our numerical results with the analytical solution (\ref{exactsolu}).
The computational domain is $[-1000,1000]\times [-250,250]\text{~m}^2$ and is partitioned with $1024\times 256$ uniform meshes. $g=9.8\text{~m}/\text{s}^2$.
Figure \ref{dambreak} depicts the comparison between the numerical solutions for two different values
of the bed friction angle and bed slope and the analytical solutions.

In figure \ref{dambreak}(a), with bed slope angle $\theta=0$ and friction angles $ \phi_\text{bed}=\phi_\text{int}=0$,  a granular mass of 20~m high, infinitely long in the positive $x$-direction on a flat bottom  is suddenly released.  It can be seen that all the three models produce the same results in good agreement with the analytical solution. In the situation of a perfectly horizontal bottom, model A is the same as model C since equation (\ref{gravity-A}) gives $g^\prime=g$ and equation (\ref{beta}) gives $\beta=1$. For model B,
since $\bar{w}$ is varying in the rarefaction zone, $D \bar{w}/ D t$  is nonzero as seen from equations (\ref{dwdt})-(\ref{wav}), causing $g^\prime \ne g$. Anyway, the numerical result has no noticeable difference from those obtained by models A and C.

In figure  \ref{dambreak}(b) with $\theta=30^\text{o}, \phi_\text{bed}=20^\text{o}$  and $ \phi_\text{int}=0^\text{o}$, a tabular reservoir of sand of $20/\cos\theta$ m high in the vertical direction on the inclined slope is released from the initial position at $x=0$ and the flow depth is shown at $t=15$~s.  It can be seen that the result of  model A is closer to the analytical solution than models B and C. Particularly,
the avalanche motion predicted by model C is the quickest and deviates most from the analytical solution.

The reason why model C gives quicker avalanche can be  explained as follows.
From the momentum component of equation (\ref{nonconserve}), the net driving force is
\begin{equation}
F_x=-g^\prime (\tan\theta -\tan\phi_\text{bed})-g^\prime \frac{\partial h}{\partial x}.
\end{equation}
Since $(\tan \theta -\tan\phi_\text{bed})>0 $ and  $ \partial h/\partial x >0$ in the whole domain, it is evident that larger $g^\prime$ has larger driving force to the negative $x$ direction, making the sand collapse faster. Thus, for model C, since  $g^\prime = g >g\cos^2\theta$, the computed avalanche flow will be faster than the analytical solution.

For models A and B, since $g^\prime $ depends on the solution, it is difficult to analyze the motion generally. However, model A can be analyzed here. Equation (\ref{gravity-A}) for this problem gives
\begin{equation}
g^\prime=\left( \frac{1+ \tan \theta   \tan \phi_\text{bed} }{\displaystyle 1+ \tan \theta \cos^2\theta \frac{\partial h}{\partial x}  }  \right)  g\cos^2\theta.
\end{equation}
Depending on whether $\cos^2\theta   (\partial h/\partial x )\gtrless \tan \phi_\text{bed}$ in different locations, $g^\prime$ can be smaller or larger than $g\cos^2\theta$, which can  make the computed avalanche motion lag behind or precede  the analytical solution in different locations as shown in figure  \ref{dambreak}(b).

\subsection {One-dimensional granular avalanche over simplified topography}

\begin{figure}[h!]
 \noindent\includegraphics[width=20pc]{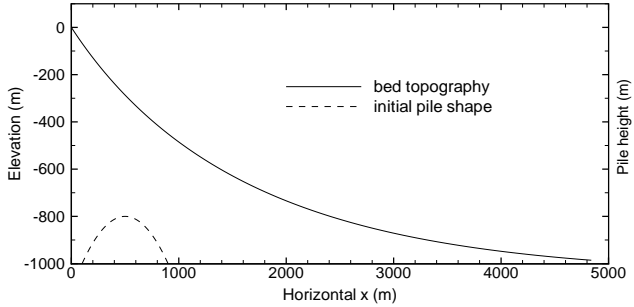}
 \caption{
 Bed topography in the horizontal Cartesian coordinates
($x,z$) and initial shape of the granular mass in the
topography-linked coordinates ($\tilde{x},\tilde{z}$).
}
\label{simpletopo}
\end{figure}

The granular avalanches over a simple transversally uniform 2-D topography \cite{Mangey2003} is chosen here to illustrate the performances of various models.
The elevation of this topography decreases from  $b=0$~m at $\tilde{x}=x=0$~m in the left end with a maximum slope inclination of 35$^\text{o}$  to $b\approx-985.9$~m at $\tilde{x}=5000$~m in the right end with slope inclination of about 2 degrees, where $\tilde{x}$ is a bed-fitted coordinate tangential to the basal surface, and $x$ is the global horizontal Cartesian coordinate. The corresponding bed slope angle and curvature are defined by
\begin{equation}
\label{simple topoangle}
\theta(\tilde{x})=\theta_0\exp\left(-\frac{\tilde{x}}{a}\right), \quad \kappa =-\frac{\text{d} \theta }{\text{d}\tilde{x} }=\frac{\theta(\tilde{x})}{a},
\end{equation}
with $\theta_0=35^\text{o}, a=1750$~m.   The topography shape $z_b=b(x)$ is parameterized with the local coordinate $\tilde{x}$ via following relations
\begin{equation}
\frac{d b}{\text{d} \tilde{x}} =-\sin \left( \theta (\tilde{x})\right), \quad
\frac{d x}{\text{d} \tilde{x}} =\cos \left( \theta (\tilde{x})\right).
\end{equation}
With the starting point chosen as $b=0, x=0$ at $\tilde{x} =0$, the topography shape can be integrated  numerically. The solid curve in figure \ref{simpletopo} shows the bed topography in the global Cartesian coordinates ($x,z$).
The dashed curve in figure \ref{simpletopo} depicts the initial parabolic shape of a pile over the
topography represented in the bed-fitted coordinate. The initial conditions of the flow are defined by the instantaneous release of the granular mass,
\begin{equation}\label{initialsimpto}
\begin{split}
&\tilde{h}(\tilde{x},t=0) =K\left[l-(\tilde{x}-\tilde{x}_0)^2\right],    \\
&\tilde{u}(\tilde{x},t=0) =0  ,
\end{split}
\end{equation}
where $K=1.25\times 10^{-3}~\text{~m}^{-1}$, $l =1.6\times 10^5\text{~m}^2$, and $\tilde{x}_0=500\text{~m}$.
Initially, the maximal thickness of the mass is 200 m in the basal normal direction with a length of 800 m in the tangential direction. In the horizontal Cartesian coordinate system,  the initial shape is the same parabolic shape centered at the projected position of $\tilde{x}_0$ and imposed on the topography in the vertical $z$ direction.

We simulate this problem by using Cartesian models A, B, and C  and
the SH model in the bed-fitted curvilinear coordinate system  \cite{Wang2004}. The
solution domain is  $[0,5000]\times [0,1250]\text{~m}^2$ for the bed-fitted model, and  $[-100,4840]\times [0,1235]\text{~m}^2$ for  the Cartesian models. The computational meshes used have the same 512 cells in the streamwise direction and 128 cells in transverse direction in both coordinate systems.

\begin{figure}[hp!]
 \noindent\includegraphics[width=20pc]{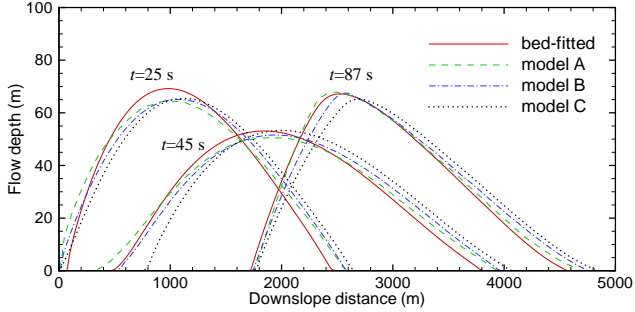}
 \caption{
 Flow depth $\tilde{h}$ vs. $\tilde{x}$ on a simplified topography at $t = 25$~s, $45$~s, and $87$~s ($t=87$~s is when the granular mass stops for model A) computed by using various models for  constant friction angles of $\phi_\text{int}=\phi_\text{bed}=15^\text{o}$.  The downslope distance is measured along the $\tilde{x}$ direction on the topography. The flow depth $\tilde{h}$ in the bed normal direction in models A, B, and C is approximated with equation (\ref{hrela}). Meanings of models A, B and C are explained in the beginning of section \ref{sec:vv}.
}
\label{simpletopo-result}
\end{figure}

Figure \ref{simpletopo-result} shows comparison of the calculated results between global Cartesian and bed-fitted models with $\phi_\text{int}=\phi_\text{bed}=15^\text{o}$. Various models produced different results of which models A and B are in better agreement  with the bed-fitted results than model C, while model C gives the fastest avalanches. The granular mass completely stops at $t = 87$ s, 85 s, and 82 s for models A, B and C respectively, and at $ t = 86.5$ s for the bed-fitted model. The maximum depth of the final deposit for model A is $\tilde{h}_{\max}= 68.4$ m, which is close to $\tilde{h}_{\max} = 67.3 $ m for the bed-fitted model. These data are also  close to those ($t_\text{stop}=86$ s and $\tilde{h}_{\max}=68$ m) calculated in Ref. \cite{Mangey2003} using a topography-linked coordinate model. However, as different definitions in ``flow depth" and ``depth-averaged velocity" exist between the global Cartesian and the bed-fitted models,  these intrinsic differences lead to different equations, thus quantitative differences between the global Cartesian and the bed-fitted models are expectable.  In general, the downstream flow front predicted by the Cartesian models propagates faster than the bed-fitted model.

\subsection {Avalanche over an inclined plane merging continuously into a horizontal plane}

\begin{figure}[hp!]
\begin{minipage}[t]{0.99\textwidth}
\includegraphics[width=20 pc]{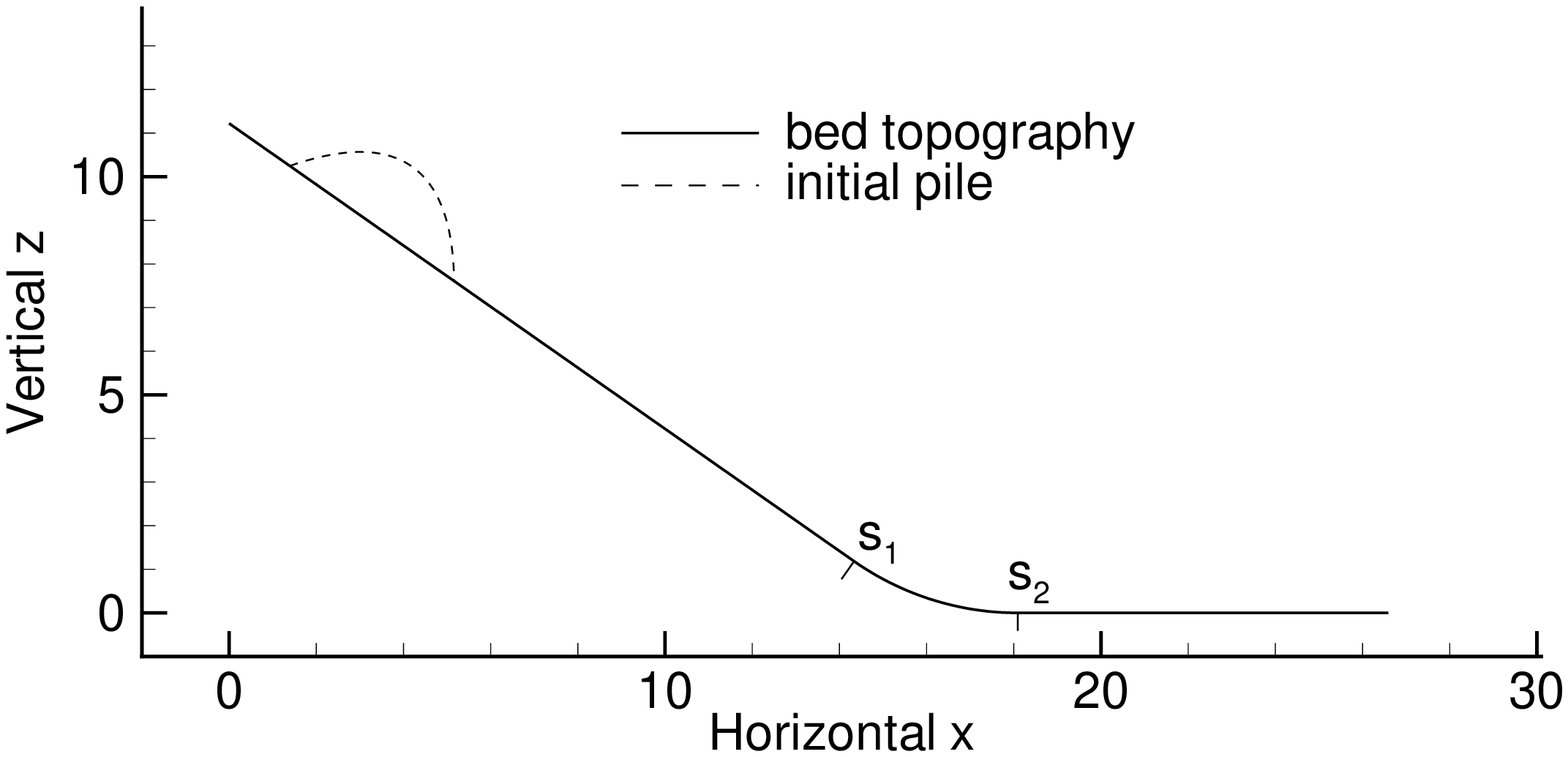}
\centering  {  \parbox{0.6\textwidth }{(a) Slope and initial pile}}
\end{minipage}%
\vskip 0.2 cm
\makebox[-0.0 cm]{}
\begin{minipage}[t]{0.99\textwidth}
\includegraphics[width=29pc]{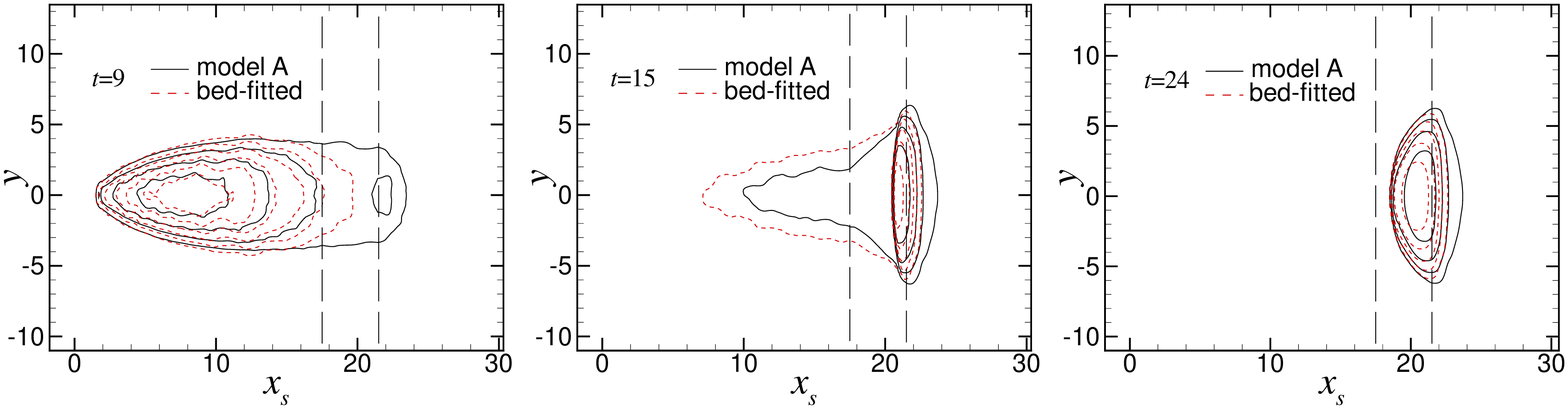}
\centering  {  \parbox{0.6\textwidth }{(b) Model A} }
\end{minipage}%
\vskip 0.2 cm
\makebox[-0.0 cm]{}
\begin{minipage}[t]{0.99\textwidth}
\includegraphics[width=29pc]{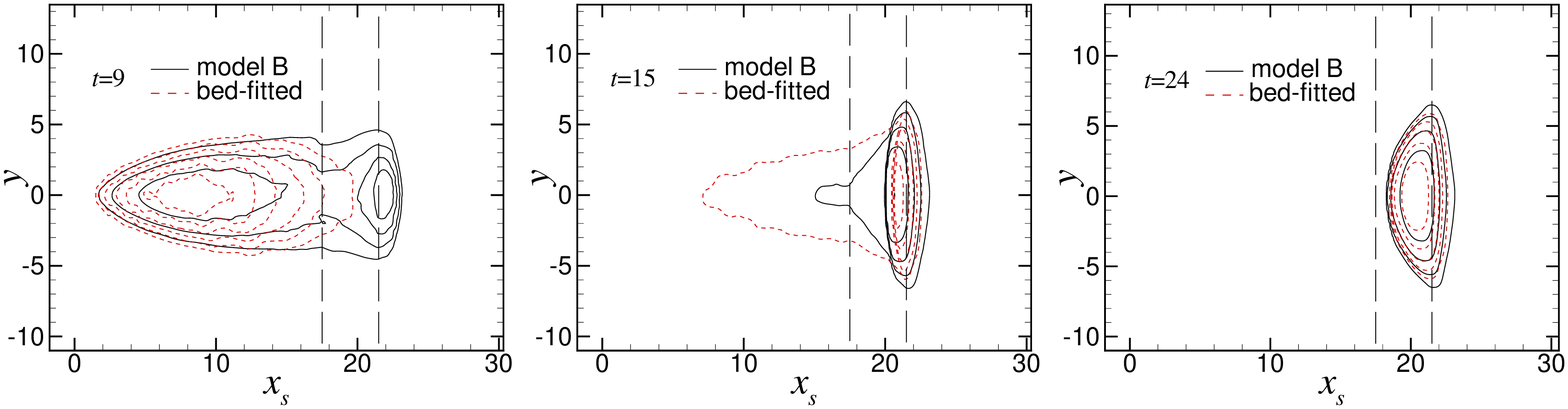}
\centering  {  \parbox{0.60\textwidth }{(c) Model B } }
\end{minipage}%
\vskip 0.2 cm
\makebox[-0.0 cm]{}
\begin{minipage}[t]{0.99\textwidth}
\includegraphics[width=29pc]{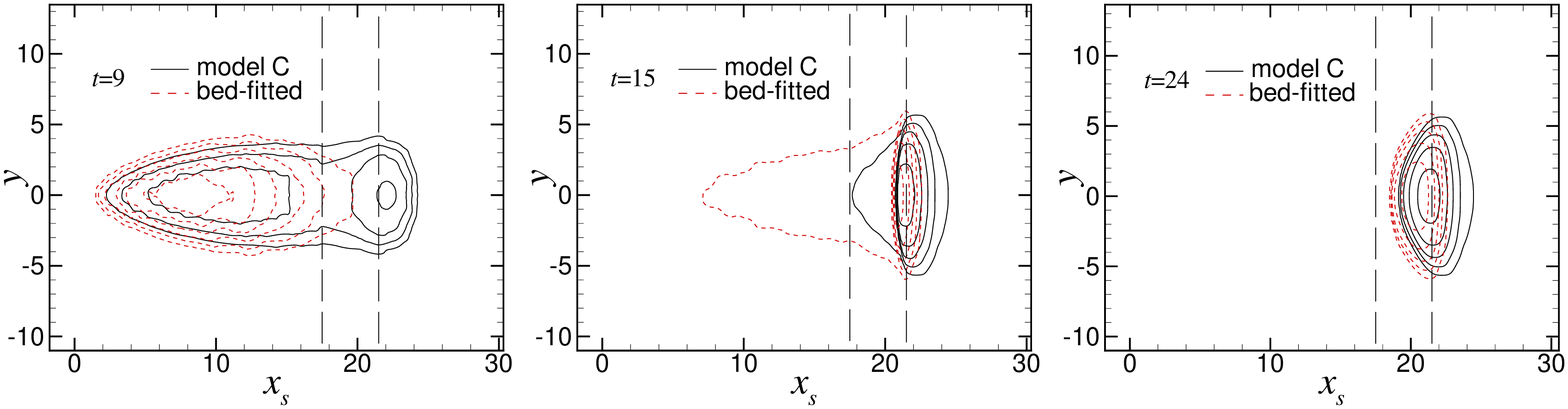}
\centering  {  \parbox{0.6\textwidth }{(d) Model C } }
\end{minipage}%
\caption{(a) Side view of the bed topography and initial pile.   The transition zone from the
inclined plane to the horizontal plane lies for $S_1=17.5 <x_s < S_2=21.5$, where $x_s$ is along the downslope direction.  (b), (c), (d) Comparison of avalanche thickness contours at times $t=9, 15 $ and $24$ computed with three Cartesian models and the bed-fitted model \cite{Wang2004}. Five equal contours from $h=0.05$ to respective maximal
depth in each frame are displayed.  The zone between two long dashed lines is the transition zone. Meanings of models A, B and C are given in the beginning of section \ref{sec:vv}.
}\label{wang}
\end{figure}

In this subsection we present a simulation example of an avalanche of finite granular mass sliding down an inclined plane and merging continuously into a horizontal plane \cite{{Wang2004}}. The problem scales are non-dimensional.
A paraboloid of rotation holding the material together is suddenly released
so that the bulk material commences to slide on an inclined flat plane at $35^\text{o}$ into a horizontal run-out plane connected by a smooth transition. For the simulation using the body-fitted coordinates $(\tilde{x},\tilde{y})$,
the computational domain is the rectangle $\tilde{x} \in [0, 30]$ and $\tilde{y} \in [-7,7]$ in dimensionless length units, where the inclined section lies in the interval $\tilde{x} \in  [0,17.5]$ and the horizontal section lies where $\tilde{x} \geq  21.5 $ with a smooth change in the
topography in the transition zone, $ \tilde{x} \in [17.5, 21.5]$. The inclination angle is prescribed by
\begin{equation}\label{angle}
  \zeta(\tilde{x})=\left\{
              \begin{array}{ll}
                \zeta_0, & 0\leq \tilde{x}\leq 17.5 ,\\
                \zeta_0 \displaystyle \left(1-\frac{\tilde{x}-17.5}{4}\right), & 17.5<\tilde{x}<21.5, \\
                0, & \tilde{x}\geq 21.5 ,\\
              \end{array}
           \right.
\end{equation}
where $\zeta_0=35^\text{o}$. The friction angles $\phi_\text{bed}=\phi_\text{int}=30^\text{o}$. A paraboloid of rotation
with height of $h_0=1.60$ and radius of $r_0=2.3$ centered at
$(\tilde{x}_0,\tilde{y}_0) = (4,0) $ is released suddenly at $t = 0$, see figure \ref{wang} (a). The initial vertical height
in the horizontal coordinates can be calculated by rotation of coordinates around the center $(\tilde{x}_0,\tilde{y}_0) = (4,0) $ with angle $\zeta_0$.

Figures \ref{wang}(b)-(d) illustrate comparison of the thickness contours of the avalanche body at three time instants ($t=9, 15$ and 24) as the avalanche slides on the inclined plane into the horizontal run-out zone. The results obtained by using different global Cartesian  models are compared with those obtained by using the bed-fitted model \cite{Wang2004}.  Comparing figures \ref{wang} (b), (c) and (d), it is seen that the avalanche speed increases from model A to model C,
and all the Cartesian models produce quicker avalanche than the bed-fitted model.
This  difference may be attributed to intrinsic differences in models such as different depth-averaging directions, as explained in the end of last subsection. The results of model A are in better agreement with the bed-fitted results. It is observed that a shock wave develops just upstream of $x_s=21.5$ at $t = 15$. With the arrival of mass from the tail, the
shock wave propagates backwards. At $t=15$, the position of the shock in model A is almost coincident with that of the bed-fitted model, the shock in model B is more upstream, while the shock in model C is more downstream probably due to shock forming at more downstream position. At $t=24$, the shock front almost reaches the beginning of the transition zone at $x_s = 17.5$ for the A, B and  bed-fitted models, and the final depositions of them  are comparable.
However, the deposition in model C is more downstream than that in the bed-fitted model.

\subsection{Granular avalanches in a chute with shallow lateral curvature}

This example was taken from Wieland \emph{et al}. \cite{Wieland1999} on the rapid fluid-like flow of a finite mass of granular material down a chute with partial lateral confinement.
The chute consists of a section inclined at 40$^{\circ}$ to the horizontal, which is connected to a plane run-out zone by a smooth transition. The reference surface is defined by the variation of its inclination angle, $\zeta$, with the
downslope coordinate $x$. The inclination angle of the reference plane is prescribed by
\begin{equation}\label{angle2}
  \zeta(x)=\left\{
              \begin{array}{ll}
                \zeta_0, &~~ x  < x_a,   \\
                \zeta_0 \displaystyle  \frac{x_b-x}{x_b-x_a}   , &~~ x_a \leq    x \leq x_b ,  \\
                0, & ~~ x_b <x  .
              \end{array}
           \right.
\end{equation}
where $\zeta_0=40^\circ$, and $x_a=175$ cm is the beginning of the
transition zone and $x_b=215$ cm is the end of the transition region. The side view of the reference plane is similar to the slope in figure \ref{wang}(a).

As described in \cite{Wieland1999}, the three dimensional basal topography is superposed
normal to the reference surface. A shallow parabolic cross-slope profile with radius of curvature $R=110$ cm  is
prescribed on the inclined section of the chute, $x<x_a=175$ cm. It opens out into a
flat run-out  zone in the region, $x>x_b=215$ cm,  and in the transition zone, $x_a\leq x \leq x_b$,  a
continuous differentiable function is constructed to provide a smooth change in the
topography. The function of the chute topography  above the reference plane, $b(x,y)$, is
\begin{equation}
b(x,y)=\left\{ \begin{array}{ll}
\displaystyle\frac{y^2}{2R},  & ~~  x<x_a, \\
\displaystyle \frac{y^2}{2R}\left[3 \left(\frac {x_b-x}{x_b-x_a}\right)^2-2\left(\frac {x_b-x}{x_b-x_a}\right)^3\right],
 &  ~~ x_a \leq x \leq x_b, \\
0 ,  & ~~  x_b<x  .
\end{array} \right.
\end{equation}

The initial condition  of the flow is the granular material packed in a hemispherical cap centered at $(x_0,y_0, z_0)=(6, 0, -(r-h_c))$ cm, which is on the parabolic cross-slope basal topography. Here, $r$ is the radius of the hemisphere and $h_c$  is the maximum height of the initial free surface  above the chute. The initial free
surface, $s(x,y)$,  of the granular material is described in the curvilinear reference coordinate system as
\begin{equation}
s(x,y) = \sqrt{r^2-x^2-y^2}-(r-h_c).
\end{equation}
The projection of the intersection of the pile edge with the basal topology onto the $z=0$ plane is approximately elliptical in shape. The major axis of the cap $r_b= 32$ cm, and the maximum height, $h_c=22$ cm. The radius $r$ is then determined by the relation $r^2=r_b^2+(r-h_c)^2$.   The pile is released from rest.

We simulate experiment V05 in Ref. \cite{Wieland1999}. The granular material is plastic beads, and we use the same fixed basal angle of friction $\phi_\text{bed}=27^\circ$ and  internal angle of friction
$\phi_\text{int}= 33^\circ$ as given in \cite{Wieland1999}.
The computational domain is $[-50, 400]\times [-70,70]$ cm$^2$ in the reference plane for the ``bed-fitted" model  computation, and is slightly extended in the horizontal $x$ direction for the Cartesian model computation.
A grid with $256\times 96$ mesh cells is used in both coordinates.

\begin{figure}[hp!]
\begin{minipage}[t]{1.0\textwidth}
\includegraphics[width=1.0\textwidth  ]{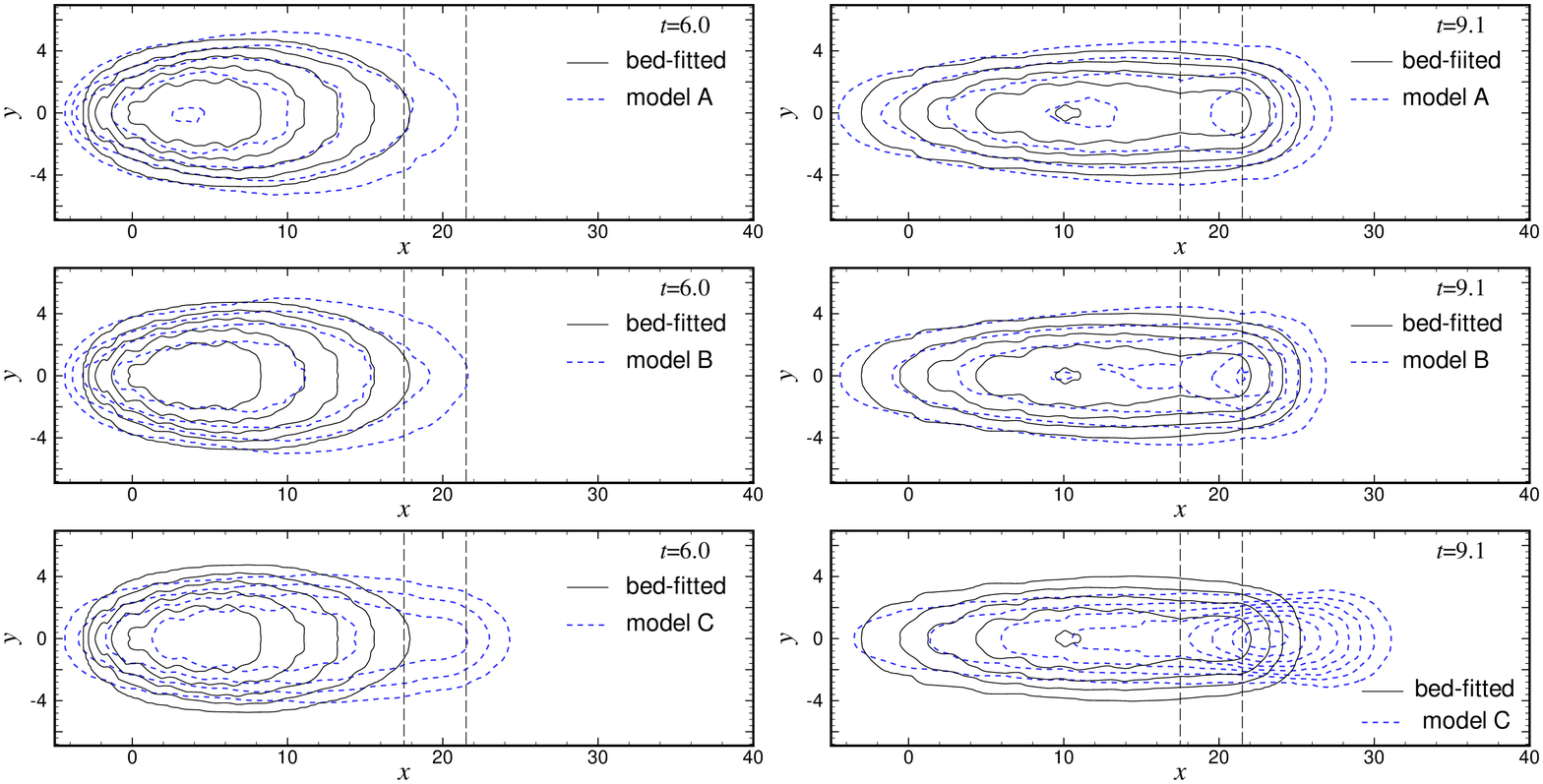}
\centering  { \parbox{0.80\textwidth }{(a) Present results computed with $k_\text{ap}=1$ at $t=6$ and $t=9.1$}}
\end{minipage}\par
\vskip 0.2cm
\begin{minipage}[t]{1.0\textwidth}
\hskip 0.0 cm
\includegraphics[width=0.49\textwidth  ]{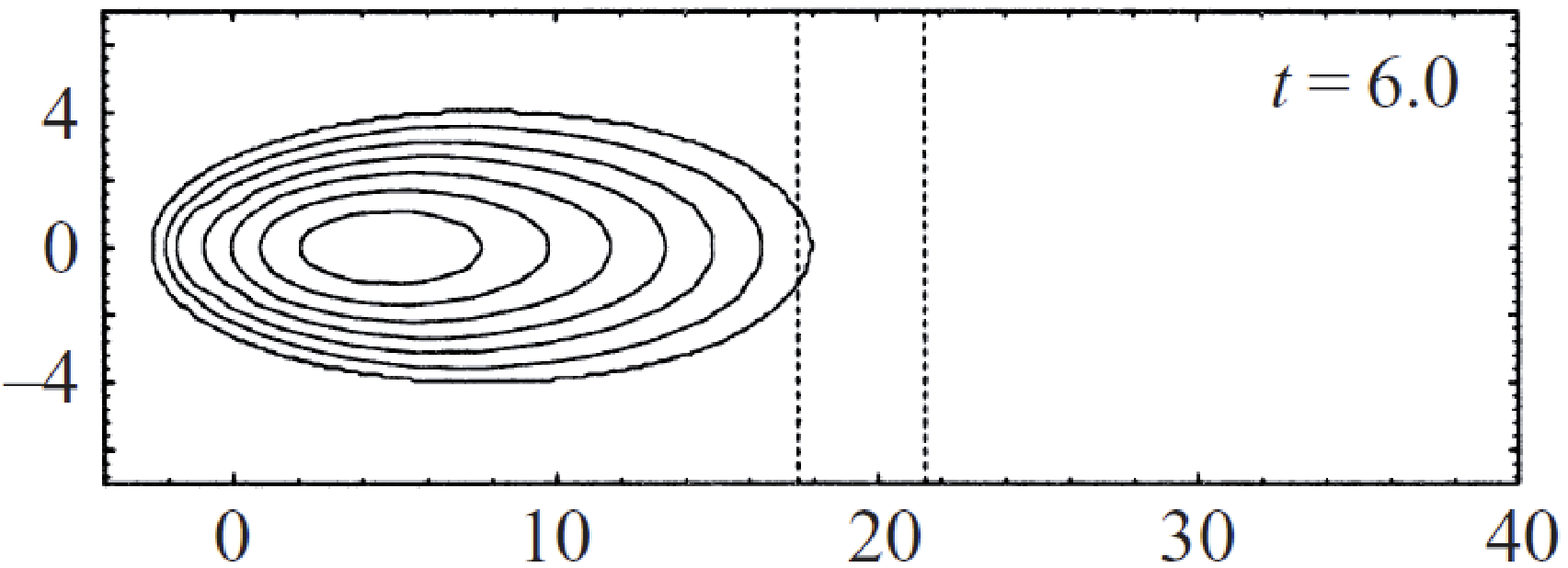} \makebox[0.0cm]{}
\includegraphics[width=0.49\textwidth  ]{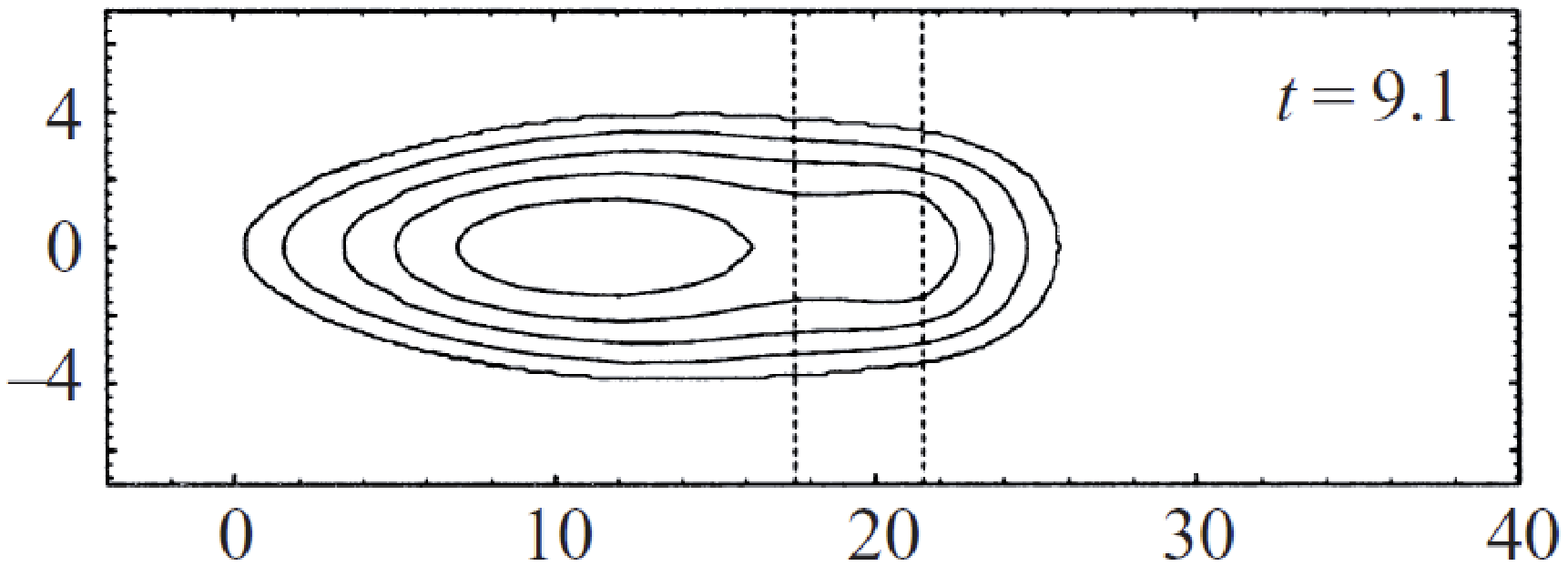}
\centering  { \parbox{0.80\textwidth }{(b) {Numerical results computed by}  Wieland \emph{et al}. \cite{Wieland1999} at $t=6$ and $t=9.1$ }}
\end{minipage}\par
\vskip 0.2cm
\begin{minipage}[t]{1.0\textwidth}
\includegraphics[width=1.0\textwidth  ]{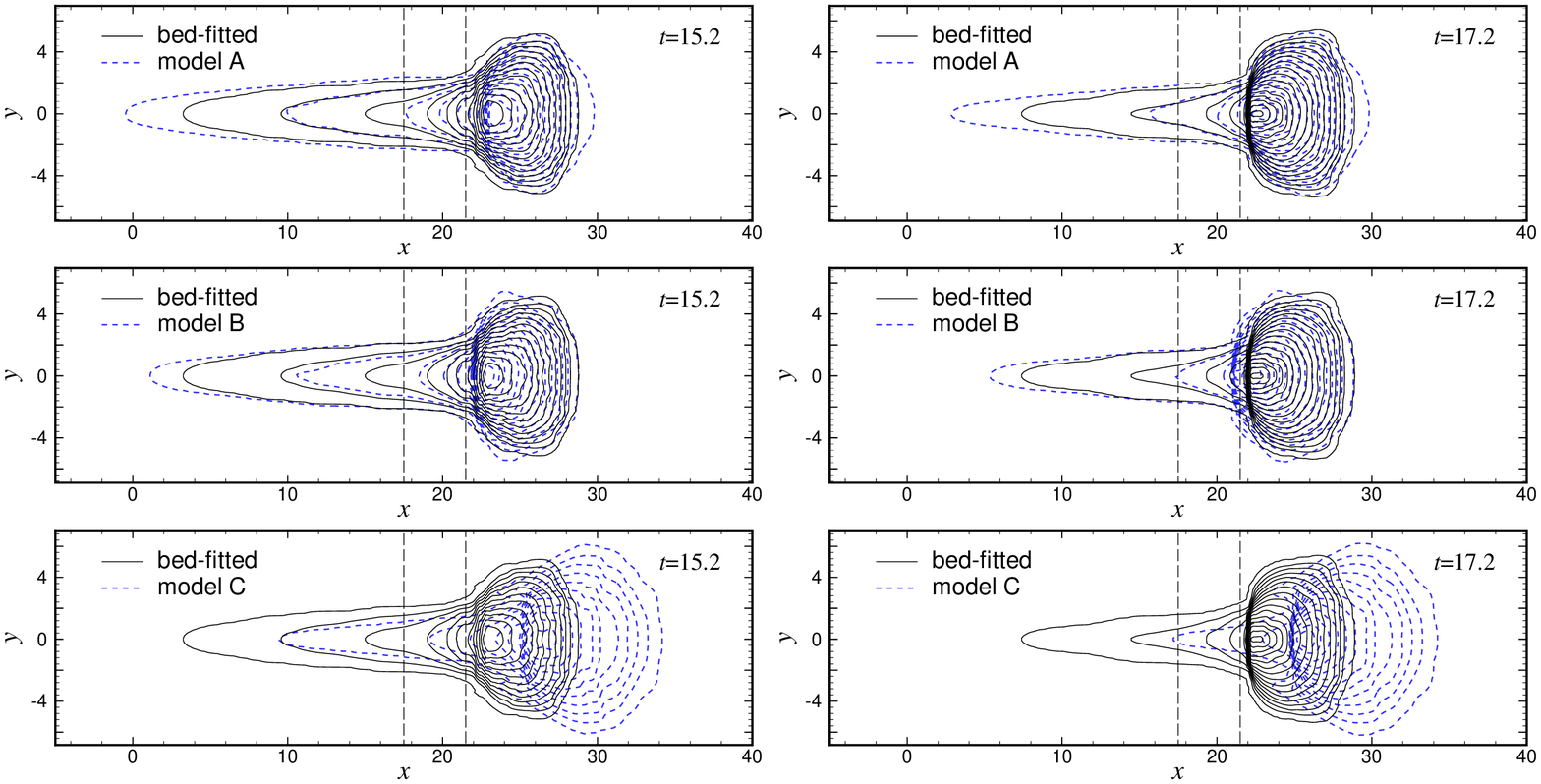}
\centering  { \parbox{0.80\textwidth }{(c) Present results computed with $k_\text{ap}=1$ at $t=15.2$ and $t=17.2$ \quad \quad}}
\end{minipage}\par
\vskip 0.2cm
\begin{minipage}[t]{1.0\textwidth}
\hskip 0.0 cm
\includegraphics[width=0.49 \textwidth  ]{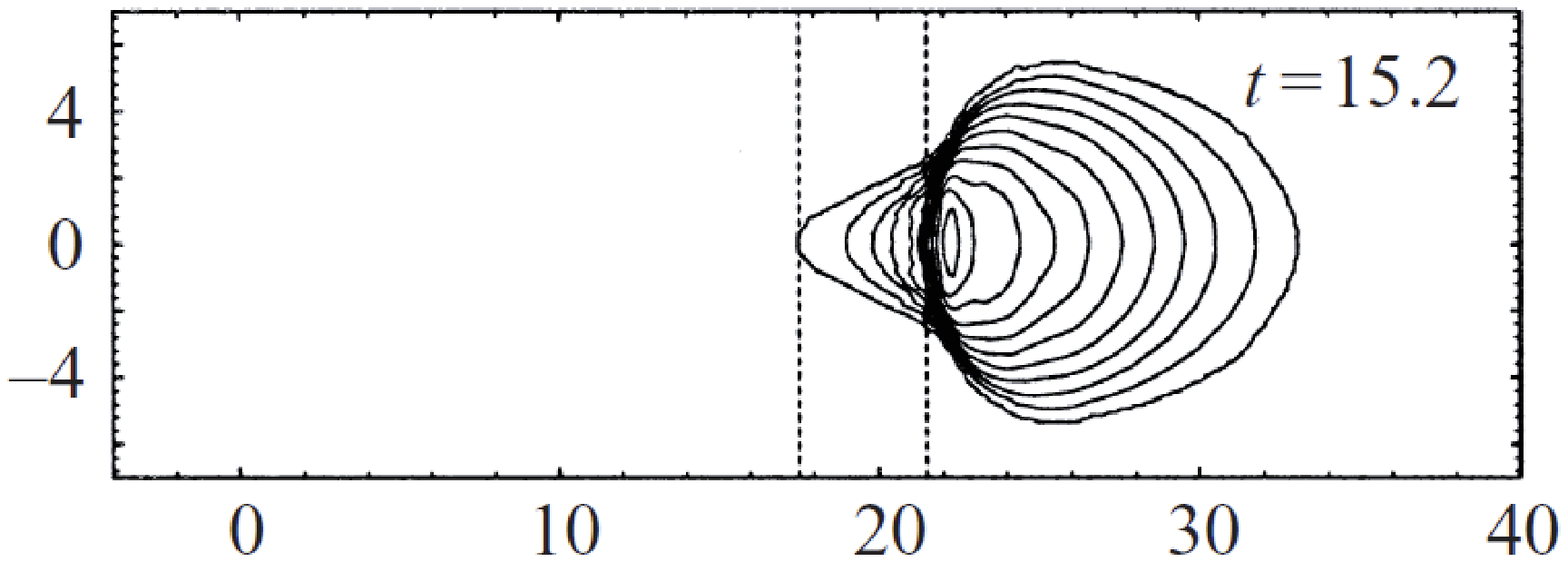} \makebox[-0.0cm]{}
\includegraphics[width=0.494 \textwidth  ]{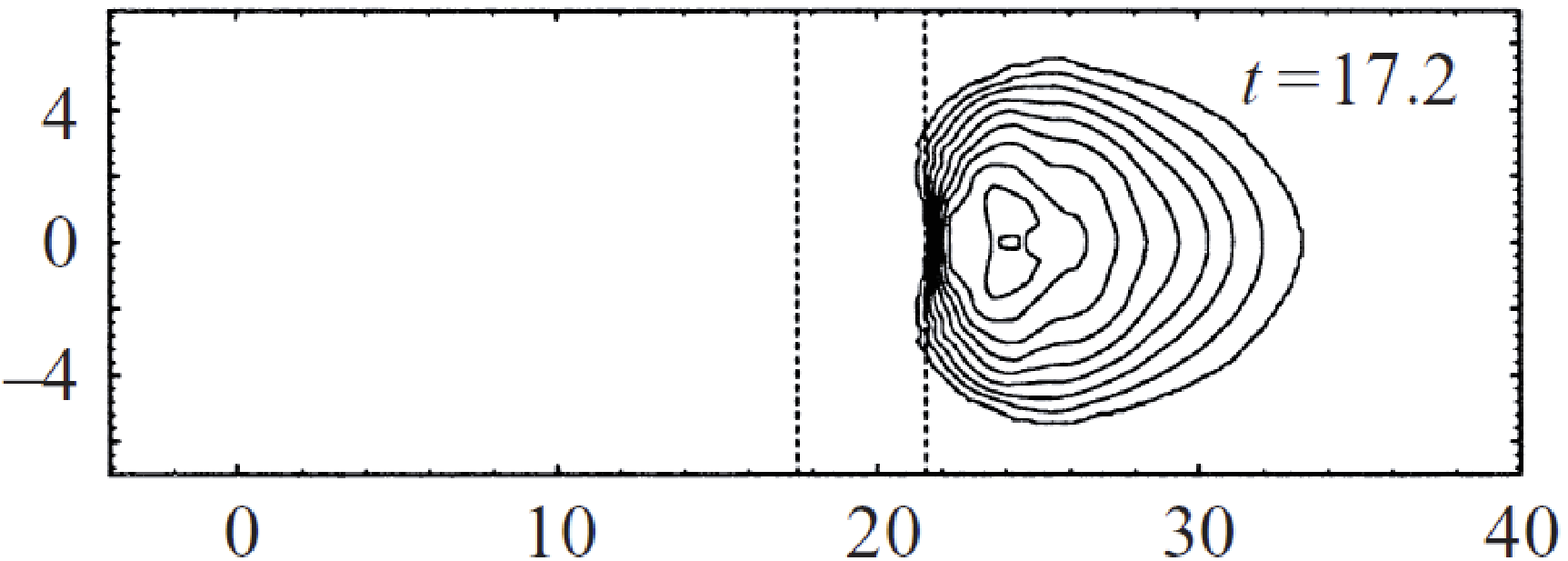}
\centering  { \parbox{0.85\textwidth }{(d) {Numerical results computed by} Wieland \emph{et al}. \cite{Wieland1999} at $t=15.2$ and $t=17.2$  }}
\end{minipage}\par
\vskip 0.2cm
\begin{minipage}[t]{1.0\textwidth}
\hskip 0.0 cm
\includegraphics[width=0.49 \textwidth  ]{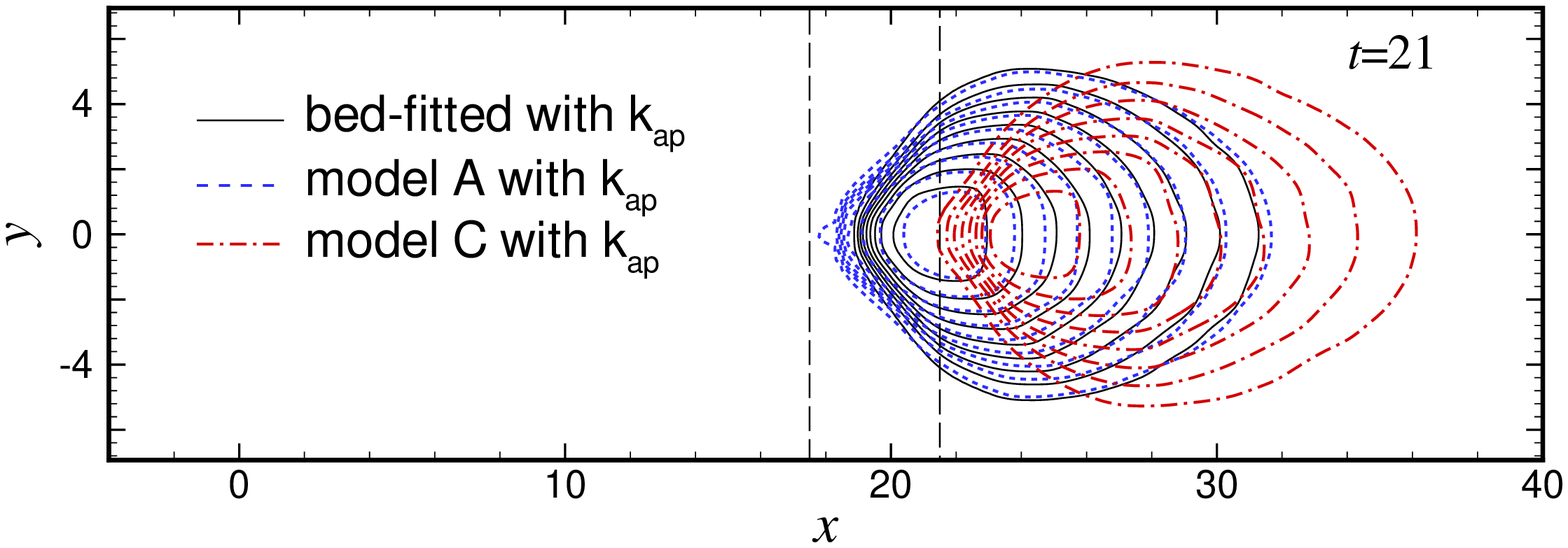}
\includegraphics[width=0.494 \textwidth  ]{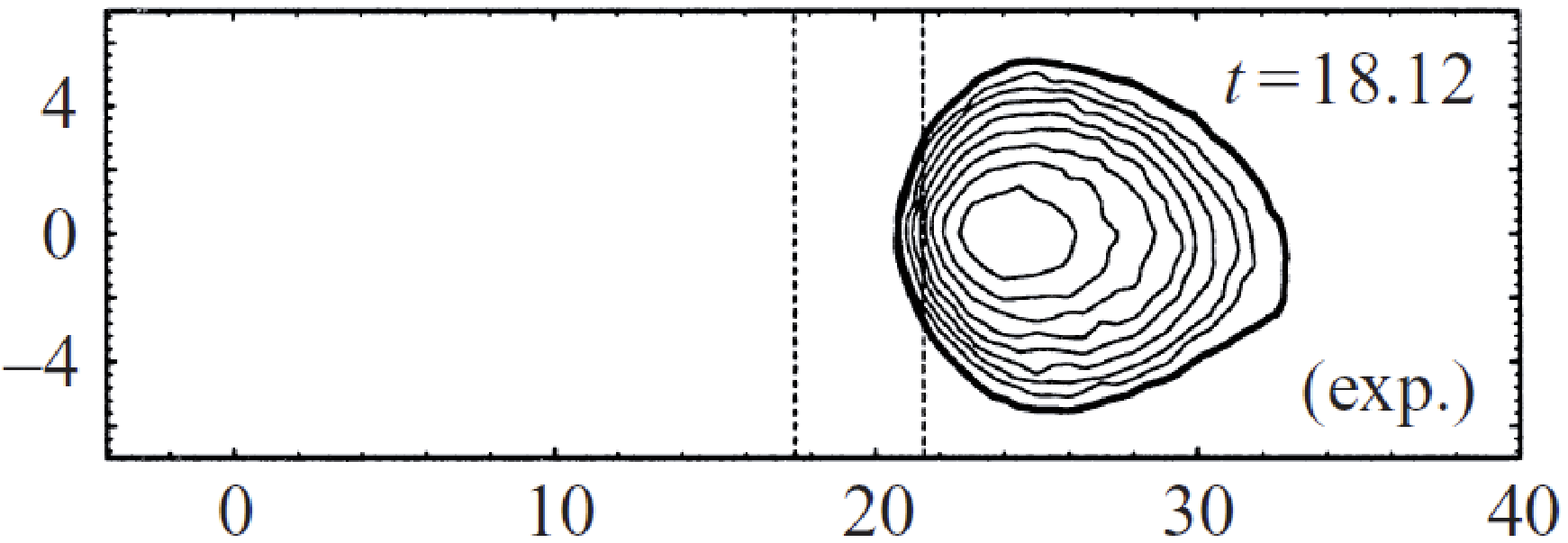}
\centering{ \parbox{0.8\textwidth }{(e) Present results (left) and experimental result of Wieland \emph{et al.}~\cite{Wieland1999} (right) }}
\end{minipage}
\caption{Comparison of present dimensionless avalanche thickness with numerical and experimental results for experiment ``V05" in Ref. \cite{Wieland1999}. {The ``bed-fitted" results in the present simulation are obtained by solving the conservative governing equations \cite{Wang2004} (equivalent to the non-conservative form \cite{Wieland1999}) with the present finite volume scheme}.   The contour levels start from pile edge (defined as 0.1~mm in present results, equivalent to $10^{-3}$ dimensionless unit) with 0.1 unit intervals. The zone between two long dashed lines is the transition zone. Meanings of models A, B and C are given in the beginning of section \ref{sec:vv}. The earth pressure coefficients  \cite{Savage1989} $k_\text{ap} \ne 1 $ are used in panel (e)-left.   }
\label{figwieland}
\end{figure}

Figure \ref{figwieland} illustrates comparisons of the computed thickness at several dimensionless times with the results  \cite{Wieland1999}. The $x$ and $y$ coordinates in the reference plane are nondimensionalized with $L_\text{ref}=10$ cm, and 0.1 unit intervals in thickness equal to 1 cm.  The ``bed-fitted" model we used is {the conservative equations written in the orthogonal curvilinear coordinate system on the reference plane \cite{Wang2004} (equation \ref{angle2})) that are reformulated from the Lagrangian form \cite{Wieland1999}}, and the equations are solved with the present finite volume scheme.
Comparing figures \ref{figwieland}(a) and \ref{figwieland}(b),  we see that the present bed-fitted results at two instants are close to the Lagrangian numerical results \cite{Wieland1999}. But all the three Cartesian models produce quicker avalanche nose and slower tail compared with the bed-fitted model, and results of models A and B are closer to the bed-fitted results than model C. Comparing figures \ref{figwieland}(c) and \ref{figwieland}(d), we can see that the shock wave in the present results begins to form at $t=15.2$, and becomes strong at $t=17.2$, while the numerical results \cite{Wieland1999} have a stronger shock wave at $t=15.2$, and it propagates upslope and becomes weak at $t=17.2$. Again, we see results of models A and B are closer to the bed-fitted results than model C.
In figure \ref{figwieland}(e), we compare our deposited avalanche thickness distributions with the final avalanche deposit in experiment V05.  For this panel we used the same earth pressure coefficients $k_\text{ap}\ne 1 $ as
given in Refs. \cite{Savage1989,Wieland1999}. It is seen that the result of model A (or B though not shown here) is very close to the  bed-fitted one, while model C predicts a deposit at a more downstream position.
The tail of the final deposit at $t=21$ computed by model A and the present ``bed-fitted" model is more upstream compared with the experimental result, yet the computed  front and span extent are comparable to the  experimental results.

\section{Conclusions}

Based on the non-hydrostatic shallow granular theory in the horizontal Cartesian system due to Castro-Orgaz \emph{et al}. \cite{Castro2014}, we {simplify the original expression of} the vertical normal stress,  and {obtain a new  formula for the basal normal stress} by using the relationship between the vertical component of the basal traction vector integrated from the $z$-momentum equation  and that of the basal Coulomb friction law.  Together with some stress relations, we turn Castro-Orgaz \emph{et al.}'s theory into a refined full non-hydrostatic shallow granular flow model in the horizontal Cartesian coordinate system. The equations are further rewritten in a form of
Boussinesq-type water wave equations {presumedly more convenient for future numerical solution using numerical methods developed in water wave field.}

For numerical solution of a low order version of the full non-hydrostatic model,  we propose an approximate formula for the enhanced gravity based on the hypothesis of hydrostatic pressure in the bed normal direction and the Taylor expansion.
In addition, we add a ``centripetal normal stress" due to the curvature tensor to the basal normal stress in the RHS terms.
The resulting simplified shallow granular flow model is implemented in the open source code  TITAN2D  for  simulating granular flows over arbitrary topography.   A series of numerical examples were carried out to test the suitability of the simplified model.  Numerical results for granular avalanches over simple topographies show that the simplified model can produce results comparable to those obtained with a topography-fitted formulation,  while the Saint-Venant equations in the horizontal Cartesian coordinates produce inaccurate results for steep slopes. It is concluded that the present simplified  model can be used to model shallow granular flows over  steep terrains.

\begin{acknowledgements}
L.~Yuan., W.~Liu, J.~Zhai thank the support of state key program for developing basic sciences (2010CB731505)
and Natural Science Foundation of China (11321061, 11261160486). S.~Wu thanks the support of Department of
Education of Guangdong Province (2014KQNCX175).
A.~Patra   and E.~Pitman acknowledge
the support of  NSF grants 0620991, 0757367 and 0711497.

The modified titan2d code can be downloaded from http:{\color{blue}//lsec.cc.ac.cn/$^{\sim}$lyuan/code.html.}
\end{acknowledgements}



\end{document}